\documentclass[11pt]{article}% добавляется twoside для двусторонней печати

\usepackage[T1]{fontenc}
\usepackage[cp1251]{inputenc}
\usepackage{textcomp}
\usepackage[centertags]{amsmath}
\usepackage[mediummath]{nccmath}
\usepackage{amsfonts}
\usepackage{amssymb}
\usepackage[pdftex]{hyperref}%hypertex,pdftex
\usepackage{subcaption}
\usepackage{graphicx}
\usepackage[numbers,sort&compress]{natbib}% упорядочивает ссылки

\usepackage{paperinitial}% Установка параметров страницы

\paperinitialization{15mm}{15mm}{15mm}{15mm}{2pt}{10pt}

\DeclareMathOperator{\sgn}{sgn}

\DeclareMathOperator{\rot}{rot}

\newcommand{\e}{\varepsilon}
\newcommand{\vf}{\varphi}

\newcommand{\al}{\alpha}
\newcommand{\be}{\beta}

\newcommand{\de}{\delta}

\newcommand{\spx}{\mathbf{x}}

\newcommand{\spk}{\mathbf{k}}

\newcommand{\spe}{\mathbf{e}}
\newcommand{\spn}{\mathbf{n}}

\begin{document}
\allowdisplaybreaks[4]% позволяет переносить многострочные формулы
\frenchspacing% уменьшение пробелов после запятых
%\sle

\title{{\Large\textbf{Transition radiation in helical metamaterials\\ with strong spatial dispersion}}}

\date{}

\author{%
P.O. Kazinski\thanks{E-mail: \texttt{kpo@phys.tsu.ru}}\;
and
P.S. Korolev\thanks{E-mail: \texttt{kizorph.d@gmail.com}}\\[0.5em]
{\normalsize Physics Faculty, Tomsk State University, Tomsk 634050, Russia}
}

\maketitle

\begin{abstract}

The theory of transition radiation in helical metamaterials with strong spatial dispersion is developed in the framework of an effective field theory approach. The average number of photons radiated by a charged particle passing through a plate made of this metamaterial is obtained. Given the positions of the transition radiation maxima in momentum space for different velocities of a charged particle, the method for reconstruction of the dispersion law of plasmon-polaritons in metamaterials is proposed. Applying this method conversely, one can predict the radiation spectrum and polarization properties of transition radiation by means of the dispersion law of plasmon-polaritons in the metamaterial known, for example, from the effective model. It is shown that the strong spatial dispersion alters qualitatively the properties of transition radiation from a charged particle traversing normally a plate made of the helical metamaterial along its symmetry axis in the paraxial regime, viz., there is a nonzero forward radiation in contrast to transition radiation in media without strong spatial dispersion. Vavilov-Cherenkov radiation and the anomalous Doppler effect in helical metamaterials with strong spatial dispersion are described.

\end{abstract}

%\slr

\section{Introduction}

%\sle

The radiation emitted by relativistic charged particles during their interaction with periodic structures represents a fundamental phenomenon underlying the operation of numerous sources of electromagnetic radiation \cite{Schwinger1949,Sokolov1968,Bazylev1987,Bagrov2002,Potylitsyn2011}. The most well-known examples are undulators \cite{Motz1951,Alferov1989}, free-electron lasers \cite{Madey1971,Freund1996,Saldin2013}, and channeling radiation \cite{Bagrov2002,BaKaStrbook,Bazylev1987,AkhShul}. The mechanisms governing the formation of this radiation encompassing transition radiation \cite{Ginzburg1945,Ginzburg1990,TerMikael1960}, diffraction radiation \cite{Bolotovskii1966,Potylitsyn2011}, Smith–Purcell radiation \cite{Smith1953}, parametric X-ray radiation \cite{Baryshevsky2005}, and Vavilov-Cherenkov (VC) radiation in periodic media \cite{Cherenkov1934,Tamm1937,Bazylev1981} are described within the general framework of quantum electrodynamics of relativistic particles in media \cite{TerMikael1973,Bazylev1987,GaribYang}. For brevity, we refer to such radiations as transition radiation \cite{Ginzburg1990}. The Bragg maxima appearing in the intensity of radiation due to periodicity of the medium are also called resonance radiation \cite{TerMikael1969}. To date, transition radiation has been thoroughly investigated for ordinary materials such as natural crystals, diffraction gratings (see, e.g., the review  \cite{Chen2023}), and plasma-like media \cite{Shroeder2004}. However, with the advent of metamaterials -- artificial media whose electromagnetic properties are determined by their subwavelength structure -- the problem of radiation from relativistic charged particles in periodic media has regained relevance \cite{konkov2025, Sergeeva2023, Sergeeva2022}. Metamaterials exhibit electromagnetic responses that are unattainable in natural materials \cite{Veselago1968, Pendry2000, Shelby2001, Seddon2003, Smith2004, Engheta2006, Duan2008, Alu2008}. Among these, the manifestation of a controllable strong spatial dispersion \cite{Landau1984, Belov2003, KK2025} is particularly important for the present study.

Alongside conventional configurations of periodical radiators, a special class of sources of transition radiation is formed by those where the periodicity exhibits a helical character. These include helical undulators, slow-wave helical structures \cite{Nezlin1976,Kompfner1947}, and helical media for generating beams of photons carrying orbital angular momentum  \cite{BKKL2021jml,BKKL2021pre,KK2022}. Of particular interest in this context are helical materials \cite{Belyakov1982, deGennes1993, Lakhtakia1995, Lakhtakia2005, Lee2005, Yang2006, Belyakov2019, Vetrov2020, KK2022} and helical metamaterials \cite{Silverinha2008, Yang2012, Gansel2012, Morgado2012_1, Morgado2012_2, Liao2014, Li2015, Jen2015, Kashke2015, Kashke2016, Morgado2016, Venkataramanababu2018}. For example, in the papers \cite{Belyakov1972, Shipov1978, Belyakov1986, Shipov1991, Velazquez2017, BKKL2021jml, BKKL2021pre}, transition radiation has been investigated in cholesteric liquid crystals \cite{Belyakov1982, deGennes1993, Yang2006} that are natural and probably the most thoroughly studied helical materials. This radiation possesses a specific spectral and angular distributions along with peculiar polarization properties \cite{Belyakov1986, Shipov1991, Velazquez2017}. Moreover, due to the helical symmetry of the medium, such radiation serves as a pure source of twisted photons \cite{BKKL2021jml, BKKL2021pre}, i.e., photons in the states carrying a nonzero projection of orbital angular momentum onto the propagation axis \cite{Rubinsztein2017, Chen2019, Padgett2017, Knyazev2018}. The presence of conducting structures combined with the chiral geometry in helical media ensures a strong coupling between the electromagnetic modes and the effective plasmonic field. The existence of the latter arises from a strong spatial dispersion that manifests in the dielectric permittivity tensor as a spatial nonlocality and a pole divergence in the momentum space.

In the recent work \cite{KK2025}, a novel effective model was proposed for describing general helical media exhibiting a strong spatial dispersion. The key idea underlying the derivation of the model consisted in expanding the dielectric permittivity tensor operator in powers of the photon momentum $\mathbf{k}$ in the vicinity of the plasmon resonance and constructing this operator from all possible tensor structures that are invariant under helical transformations. A particular case of such media is represented by helical metamaterials \cite{Silverinha2008, Yang2012, Gansel2012, Morgado2012_1, Morgado2012_2, Liao2014, Li2015, Jen2015, Kashke2015, Kashke2016, Morgado2016, Venkataramanababu2018}, i.e., composite materials consisting of an ordered array of conducting helices. The model enables the description of the electromagnetic properties of helical metamaterials at wavelengths on the order of the helix pitch, where a strong interaction between plasmonic and electromagnetic modes occurs. In particular, in this regime, polarization-dependent (chiral) band gaps are revealed that are analogous to those observed in cholesteric liquid crystals \cite{Belyakov1982, deGennes1993, Lakhtakia1995, Lakhtakia2005, Lee2005, Yang2006, Belyakov2019, Vetrov2020, KK2022}. Notice that the conventional approach to describing such media, namely, through the procedure of a complete homogenization, fails to reproduce this effect as information about the helical periodic nature of the material is lost upon averaging. The symmetry-based approach developed within the framework of an effective field theory does not suffer from this drawback and provides a clear and transparent interpretation of the electromagnetic processes occurring in such a medium. When applied to scattering problems, the proposed model demonstrated an excellent agreement with independent numerical simulations and earlier experiments \cite{Wu2010, LiDong2015}. In the particular case of the medium comprised of thin straight conducting wires, the dielectric permittivity arising in the effective model goes to the known expression derived in \cite{Belov2003}.

In the present paper, we employ this effective model to describe transition radiation from relativistic charged particles propagating in helical metamaterials with strong spatial dispersion. It turns out that the presence of a strong spatial dispersion changes drastically the properties of transition radiation and results in new effects. In particular, one of such effects is a nonzero intensity of the forward radiation from a charged particle traversing normally a plate made of the helical metamaterial along its symmetry axis. This forward radiation vanishes for dielectric helical and isotropic materials. We describe the properties of forward transition radiation in the helical metamaterials with strong spatial dispersion and show that it is caused by the existence of the plasmon degree of freedom.

The paper is organized as follows. In Sec. \ref{IntroModel}, we present the model of a helical metamaterial exhibiting a strong spatial dispersion. By explicitly introducing the plasmonic field into the theory, we eliminate the nonlocality in the dielectric permittivity tensor and formulate the Maxwell equations in a local form. Section \ref{Radiation} is devoted to the theoretical description of transition radiation, where we adapt the well-known formula for the average number of photons radiated by a classical current to the case of a charged particle moving in the helical metamaterial. In Sec. \ref{NwaveApprox}, we develop the procedure for constructing approximate solutions to the Maxwell equations in the helical metamaterial with strong spatial dispersion using the $(2N+1)$-wave approximation. This approach proves to be more efficient for obtaining solutions in a periodic medium with a large number of lattice periods \cite{Sachdev1994,Shipov2016,Vila2017,Haas2020,BKKL2021jml} then the direct solution of the system of ordinary differential equations (ODEs) stemming from the Maxwell equations. Section \ref{NumericalSimulation} presents the results of the numerical simulations of transition radiation in the helical metamaterial under various regimes. Then, in Sec. \ref{ForwardRad}, we describe a distinct feature of transition radiation in metamaterials with strong spatial dispersion -- the presence of forward transition radiation. In conclusion, we summarize the results. Throughout the paper we use the system of units such that $c=\hbar=1$ and $e^2=4\pi\alpha$, where $\alpha\approx 1/137$ is the fine-structure constant.

\section{Model of a helical medium}\label{IntroModel}

Let us introduce the model of the helical metamaterial with strong spatial dispersion developed in \cite{KK2025}. Based on the symmetry reasonings, we found the general form for the permittivity tensor  $\hat{\e}_{ij}$ with strong spatial dispersion for such a medium in the leading order in derivatives. The kernel $K_{ij}(k_0,\spx,\spx')$ of the operator $\hat{\e}_{ij}$ has the from
\begin{equation}\label{ModelKernel}
    K_{ij}(k_0,\spx,\spx')=\e_h \Big[ \de_{ij}- \tau_i(z) \frac{ \omega_p^2}{\omega_0^2
	-v^2(\boldsymbol{\tau}(z)\hat{\spk})^2}\tau_j(z')\Big]\de(\spx-\spx'),
\end{equation}
where $\omega_0:=\sqrt{\e_h}k_0$, $k_0$ is the energy of the electromagnetic wave, $\hat{k}_i=-i\partial/\partial x_i$, $\e_h$ is the dielectric permittivity of the medium where the thin conducting wires are placed, the constant $v$ is interpreted as the velocity of propagation of plasmons along the wires,
\begin{equation}
    \boldsymbol{\tau} (z)= \cos\alpha \mathbf{e}_3+\sin\alpha \mathbf{d}(z),\qquad \mathbf{d}(z)=(\cos(qz),\sin(qz),0),\qquad \alpha\in [0,\pi/2].
\end{equation}
Here $\{\spe_1,\spe_2,\spe_3\}$ are the standard unit basis vectors, $\boldsymbol{\tau}(z)$ is a unit tangent vector to the helix, $\alpha$ is a helix angle, $q$ defines the helix pitch $p:=2\pi/|q|$, and $\sgn(q)$ determines a handedness of the helix.

Since we know the expression for the permittivity, we can Fourier transform the Maxwell equations in dispersive medium with respect to time and write them as
\begin{equation}\label{MaxwellEq}
	\big(\text{rot}^2_{ij}-k_0^2\hat{\e}_{ij}\big)A_j=0,\qquad \hat{k}_i(\e_{ij}A_j)=0,
\end{equation}
where $\mathbf{A}(k_0,\spx)$ is the Fourier transformed electromagnetic potential and the second equation in \eqref{MaxwellEq} fixes the gauge. The gauge fixing condition follows from the first equation in \eqref{MaxwellEq} for $k_0\neq0$. Henceforth, we suppose that $k_0\neq0$ and, consequently, consider only the first equation.

The Maxwell equations \eqref{MaxwellEq} are equivalent to the system of equations
\begin{equation}\label{MaxwellEq0}
\begin{split}
	(\omega_0^2 - v^2 (\boldsymbol{\tau}\hat{\spk})^2) \Psi + \omega_0 \omega_p (\boldsymbol{\tau}\mathbf{A})&=0,\\
	(\omega_0^2-\text{rot}^2)\mathbf{A}+\omega_0\omega_p\Psi\boldsymbol{\tau}&=0,
\end{split}
\end{equation}
where we have got rid of nonlocality in permittivity tensor \eqref{ModelKernel} by introducing the scalar plasmon field $\Psi$ in the medium. We assume that the helical medium occupies the region of space $z\in(-L,0)$, whereas the outside of the metamaterial sample is filled by a homogeneous isotropic dielectric medium with permittivity $\e_0$. In addition to the standard boundary conditions imposed on the electromagnetic field, it is necessary to require that the plasmon field vanishes at the boundary of the metamaterial \cite{Pekar1,Pekar2}. Then the complete set of boundary conditions becomes
\begin{equation}\label{PsiBoundCond}
    [\mathbf{A}_\perp]_{z=-L}=[\mathbf{A}_\perp]_{z=0}=0,\qquad[\rot\mathbf{A}_\perp]_{z=-L}=[\rot\mathbf{A}_\perp]_{z=0}=0,
    \qquad\Psi(-L)=\Psi(0)=0.
\end{equation}

As long as the system of equations under study is translation invariant in the $(x,y)$ plane, we seek for a
solution of equations \eqref{MaxwellEq0} in the form
\begin{equation}\label{Anzatz1}
    \mathbf{A}(\mathbf{x})=e^{i\mathbf{k}_\perp \mathbf{x}_\perp}\mathbf{A}(z),\quad \Psi(\mathbf{x})=e^{i\mathbf{k}_\perp \mathbf{x}_\perp}\Psi(z).
\end{equation}
Then the Maxwell equations are reduced to the system of ODEs
\begin{equation}\label{MaxwellEq1}
\begin{split}
    \omega_0^2 a_+ -\frac{k_\perp^2}{2}(a_+-a_-)-\frac{k_\perp^2}{2\bar{k}^2_3}\hat{k}_3^2(a_++a_-)-\hat{k}_3^2a_+ +\frac{\omega_p \omega_0}{\sqrt{2}}\big[\sin\al e^{i\theta} -\frac{k_\perp\cos\al}{\bar{k}_3^2}\hat{k}_3 \big]\tilde{\Psi}&=0,\\
    \omega_0^2 a_- +\frac{k_\perp^2}{2}(a_+-a_-)-\frac{k_\perp^2}{2\bar{k}^2_3}\hat{k}_3^2(a_++a_-)-\hat{k}_3^2a_- +\frac{\omega_p \omega_0}{\sqrt{2}}\big[\sin\al e^{-i\theta} -\frac{k_\perp\cos\al}{\bar{k}_3^2}\hat{k}_3 \big]\tilde{\Psi}&=0,\\
    \big[\omega_0^2 -v^2(\cos\al\hat{k}_3 +k_\perp\sin\al\cos\theta)^2 -\frac{\cos^2\al \omega_p^2\omega_0^2}{\bar{k}_3^2} \big]\tilde{\Psi} +\frac{\omega_p\omega_0}{\sqrt{2}}\big[\sin\al (e^{i\theta} a_-+e^{-i\theta}a_+) -\frac{k_\perp\cos\al}{\bar{k}_3^2}\hat{k}_3(a_++a_-) \big]&=0,
\end{split}
\end{equation}
where the arguments of the unknown functions are omitted for brevity, and the following notation has been introduced
\begin{equation}\label{FieldNotation}
    \mathbf{A}(z)=a_3(z)\mathbf{e}_3 + \frac{1}{2}(a_{+}(z)e^{i \vf}\mathbf{e}_{-}+a_{-}(z)e^{-i \vf}\mathbf{e}_{+}),\qquad \tilde{\Psi}=\sqrt{2}\Psi,
\end{equation}
and
\begin{equation}
    \theta:=qz-\vf, \qquad \vf = \arg (k_+), \qquad \bar{k}_3:=(\omega_0^2-k_\perp^2)^{1/2}.
\end{equation}
We also use the basis vectors $\mathbf{e}_{\pm}=\mathbf{e}_1 \pm i\mathbf{e}_2$ and the components $k_\pm:=k_1\pm i k_2$. The component $A_3$ of the electromagnetic potential is expressed as
\begin{equation}\label{A_3}
    A_3(z)=a_3(z)=- \cos \alpha\frac{\omega_0\omega_p}{\bar{k}_3^2}\Psi(z) -\frac{k_\perp}{2\bar{k}_3^2}\hat{k}_3(a_+(z)+a_-(z)).
\end{equation}

The system of equations \eqref{MaxwellEq1} is periodic in the variable $z$. Therefore, we seek for a solution to this system in the form of a Fourier series
\begin{equation}\label{FourierAnzatz}
\begin{split}
    a_\pm = \sum_{l=-\infty}^\infty a_{\pm}^{l}(k_3)e^{i(k_3\pm q) \theta/q}e^{il\theta}, \qquad
    \tilde{\Psi}= \sum_{l=-\infty}^\infty \tilde{\Psi}^l(k_3) e^{ik_3\theta/q}e^{il\theta}.
\end{split}
\end{equation}
On substituting \eqref{FourierAnzatz} into \eqref{MaxwellEq1}, we arrive at the infinite system of coupled algebraic linear equations
\scriptsize
\begin{equation}\label{MaxwellEq3}
\begin{split}
    \omega_0^2 a_+^{l} -\frac{k_\perp^2}{2}(a_+^{l}-a_-^{l+2})-\frac{k_\perp^2}{2\bar{k}^2_3}(k_3+q+ql)^2(a_+^{l}+a_-^{l+2})-(k_3+q+ql)^2a_+^{l} +\frac{\omega_p \omega_0}{\sqrt{2}}\big[\sin\al \tilde{\Psi}^{l} -\frac{k_\perp\cos\al}{\bar{k}_3^2}(k_3+q+ql) \tilde{\Psi}^{l+1}\big]&=0,\\
    \omega_0^2 a_-^{l} +\frac{k_\perp^2}{2}(a_+^{l-2}-a_-^{l})-\frac{k_\perp^2}{2\bar{k}^2_3}(k_3-q+ql)^2(a_+^{l-2}+a_-^{l})-(k_3-q+ql)^2a_-^{l} +\frac{\omega_p \omega_0}{\sqrt{2}}\big[\sin\al  \tilde{\Psi}^{l} -\frac{k_\perp\cos\al}{\bar{k}_3^2}(k_3-q+ql)  \tilde{\Psi}^{l-1}\big]&=0,\\
    \big[\omega_0^2  -\frac{\cos^2\al \omega_p^2\omega_0^2}{\bar{k}_3^2} -v^2\cos^2\alpha (k_3+ql)^2-\frac{v^2k_\perp^2\sin^2\alpha}{2} \big]\tilde{\Psi}^{l} +\frac{\omega_p\omega_0}{\sqrt{2}}\big[\sin\al (a_-^{l}+a_+^{l}) -\frac{k_\perp\cos\al}{\bar{k}_3^2}(k_3+ql)(a_+^{l-1}+a_-^{l+1}) \big]&-\\
    -\frac{v^2 k_\perp^2 \sin^2\alpha}{4}(\tilde{\Psi}^{l-2}+\tilde{\Psi}^{l+2}) - \frac{v^2k_\perp \sin 2\alpha }{2} \big[(k_3+ql-\frac{q}{2})\tilde{\Psi}^{l-1} + (k_3+ql+\frac{q}{2})\tilde{\Psi}^{l+1}\big]&=0.
\end{split}
\end{equation}
\normalsize
It is easy to see that this system closes on the coefficients $a_\pm^l,\tilde{\Psi}^l$ for $k_\perp=0$ and become exactly solvable (see for details \cite{KK2025}). In the general case, the system of equations \eqref{MaxwellEq3} possesses a nontrivial solution only for six values of the quasi-momentum $k_3^{i}$, $i=\overline{1,6}$, since we have the three second order ODEs in \eqref{MaxwellEq1}. As a result, we obtain the six linearly independent solutions $\mathbf{A}^i$ and $\Psi^i$, and the general solution is written as their linear combination
\begin{equation}
	\mathbf{A}=\sum_{i=1}^{6}b_i\mathbf{A}^i,\qquad	\Psi=\sum_{i=1}^{6}b_i\Psi^i.
\end{equation}
Using the notation \eqref{FieldNotation}, the field components become
\begin{equation}\label{ModeMedium}
\begin{split}
    a_\pm(z) &= \sum_{i=1}^{6} b_i \sum_{l=-\infty}^\infty a_{\pm}^l(k_3^i) e^{i(k_3^i\pm q) \theta/q}e^{il\theta},\\
    \tilde{\Psi}(z)&= \sum_{i=1}^{6} b_i \sum_{l=-\infty}^\infty \tilde{\Psi}^{l}(k_3^i) e^{i k_3^i \theta/q}e^{il\theta},
\end{split}
\end{equation}
and $a_3(z)$ is found from the relation \eqref{A_3}. The arbitrary complex constants $b_i$ are fixed by the boundary conditions \eqref{PsiBoundCond} and the normalization condition.

\section{Electromagnetic radiation in a medium}\label{Radiation}

Let us describe the emission of photons by a classical current in a dispersive medium. The average number of photons produced by a charged particle in a medium is written as \cite{TerMikael1969,TerMikael1973,Ginzburg1979}
\begin{equation}\label{GeneralProb}
    dP(s,\spk)= Z^2 e^2 \bigg|\int_{-\infty}^{\infty}dt \dot{\mathbf{x}}\mathbf{\Phi}(s,\mathbf{k};x(t))\bigg|^2 \frac{Vd\mathbf{k}}{(2\pi)^3},
\end{equation}
where $V$ is the normalization volume and $Z$ is the charge number. The mode functions of the electromagnetic potential $\mathbf{\Phi}(s,\mathbf{k};\mathbf{x})$ matched at the interface take the form
\begin{equation}
    \mathbf{\Phi}(s,\mathbf{k};\mathbf{x})=
    \begin{cases}
    \frac{c}{\sqrt{2k_0V}} \mathbf{f}(s,\spk)e^{-ik_0x^0 +i\spk\spx},\quad &z>0;\\
    \frac{c}{\sqrt{2k_0V}}\bigg( [d_+\mathbf{f}_{++}+d_-\mathbf{f}_{-+}]e^{ik_3z}+[h_+\mathbf{f}_{+-}+h_-\mathbf{f}_{--}]e^{-ik_3z}\bigg)e^{-ik_0x^0+i\spk_\perp \spx_\perp},\quad &z<-L;\\
    \frac{c}{\sqrt{2k_0V}}\big[a_{3}(z)\mathbf{e}_3+\frac{1}{2}a_{+}(z)e^{i\vf}\mathbf{e}_{-}+\frac{1}{2}a_{-}(z)e^{-i\vf}\mathbf{e}_{+}\big]e^{-ik_0x^0+i\spk_\perp \spx_\perp}, \quad &-L \leq z\leq 0,\\
    \end{cases}
\end{equation}
where $\mathbf{f}(s,\spk)$ is the vector of a circular polarization $s$,
\begin{equation}
	\mathbf{f}(s,\spk)=(n_3 \cos\vf -is\sin\vf, n_3\sin\vf+is\cos\vf,-n_\perp )/\sqrt{2},\qquad \mathbf{n}:=\spk/k_0,
\end{equation}
and
\begin{equation}
	\mathbf{f}_{++}:=\mathbf{f}(1,\spk),\qquad 	\mathbf{f}_{-+}:=\mathbf{f}(-1,\spk),\qquad
    \mathbf{f}_{+-}:=\mathbf{f}(1,\spk_\perp,-k_3),\qquad \mathbf{f}_{--}:=\mathbf{f}(-1,\spk_\perp, -k_3).
\end{equation}
The constants $c$, $d_\pm$, $h_\pm$, and $b_i$ that enter into $a_3(z)$ and $a_\pm(x)$ are determined by the boundary conditions \eqref{PsiBoundCond}, and the normalization condition for the mode functions of a quantum electromagnetic field. The contribution to the normalization condition from a finite-width metamaterial sample is negligible as compared with the contribution from the infinite region outside of the plate \cite{BKKL2021jml,BKL2019}. Then the normalization condition reads
\begin{equation}
    |c|^2=\big(|d_+|^2+|d_-|^2\big)^{-1}.
\end{equation}

%%%%%%%%figure 1 here
\begin{figure}[tp]
	\centering
	\includegraphics*[width=0.32\linewidth]{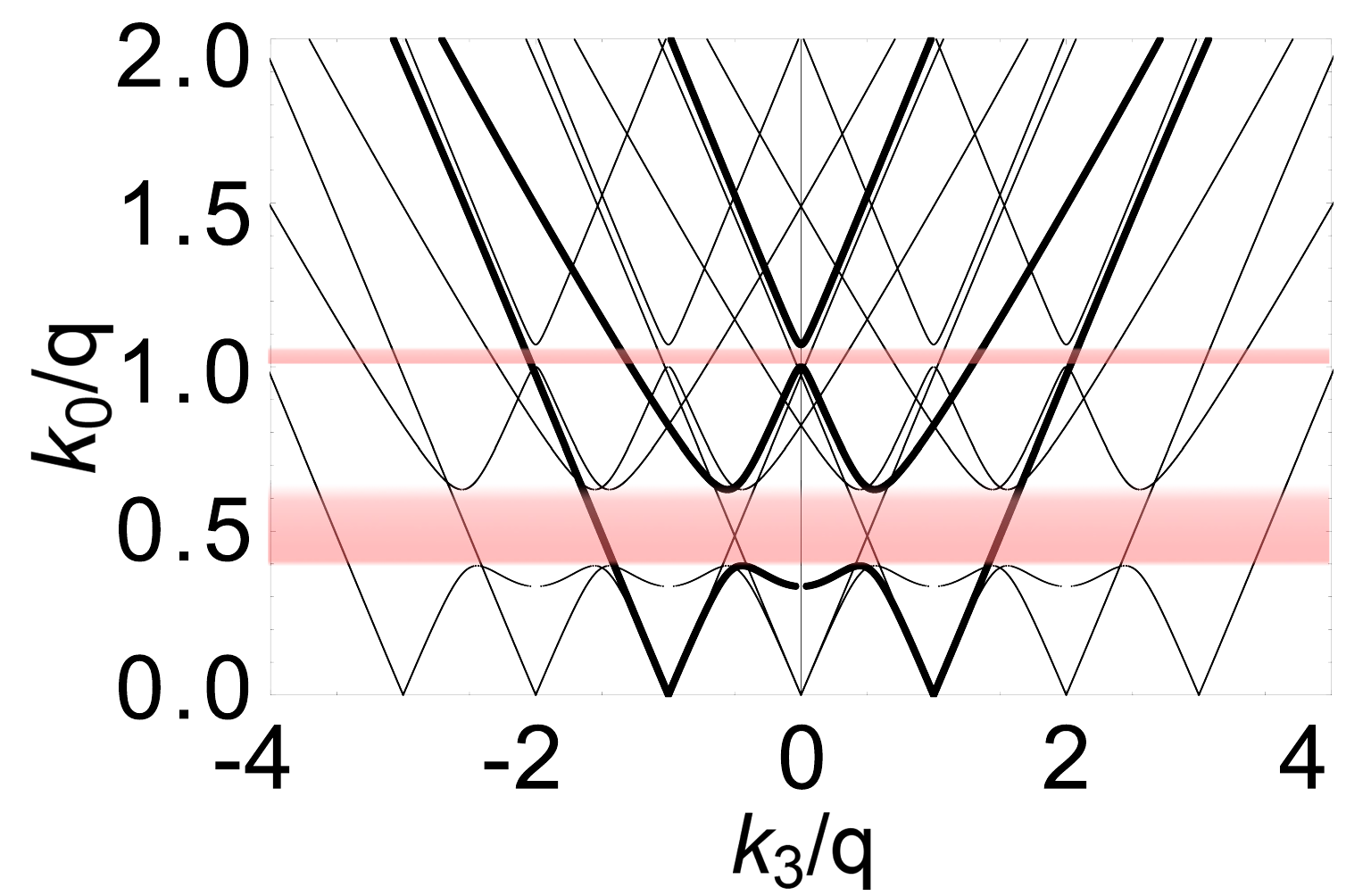}
	\includegraphics*[width=0.32\linewidth]{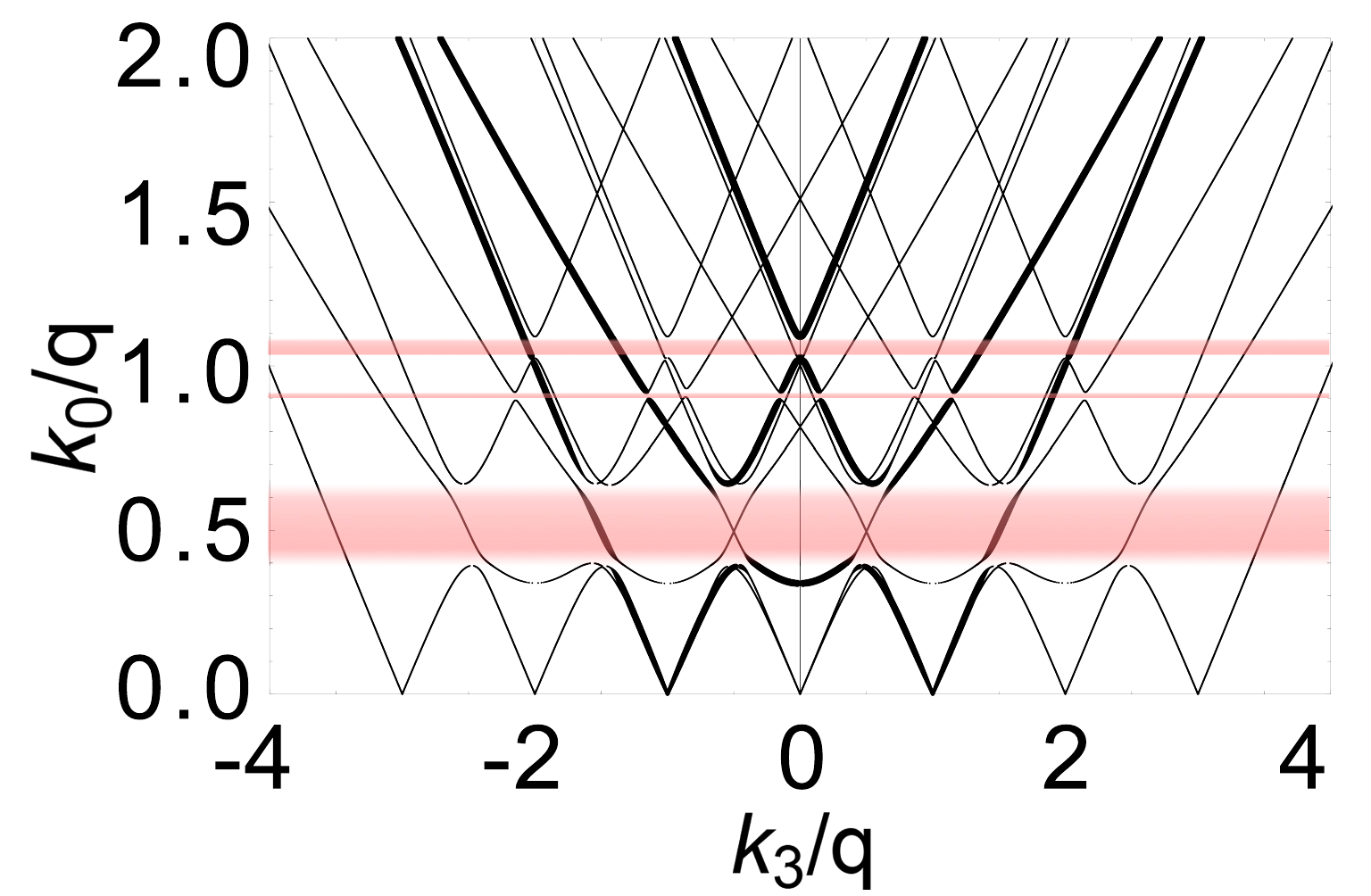}
	\includegraphics*[width=0.32\linewidth]{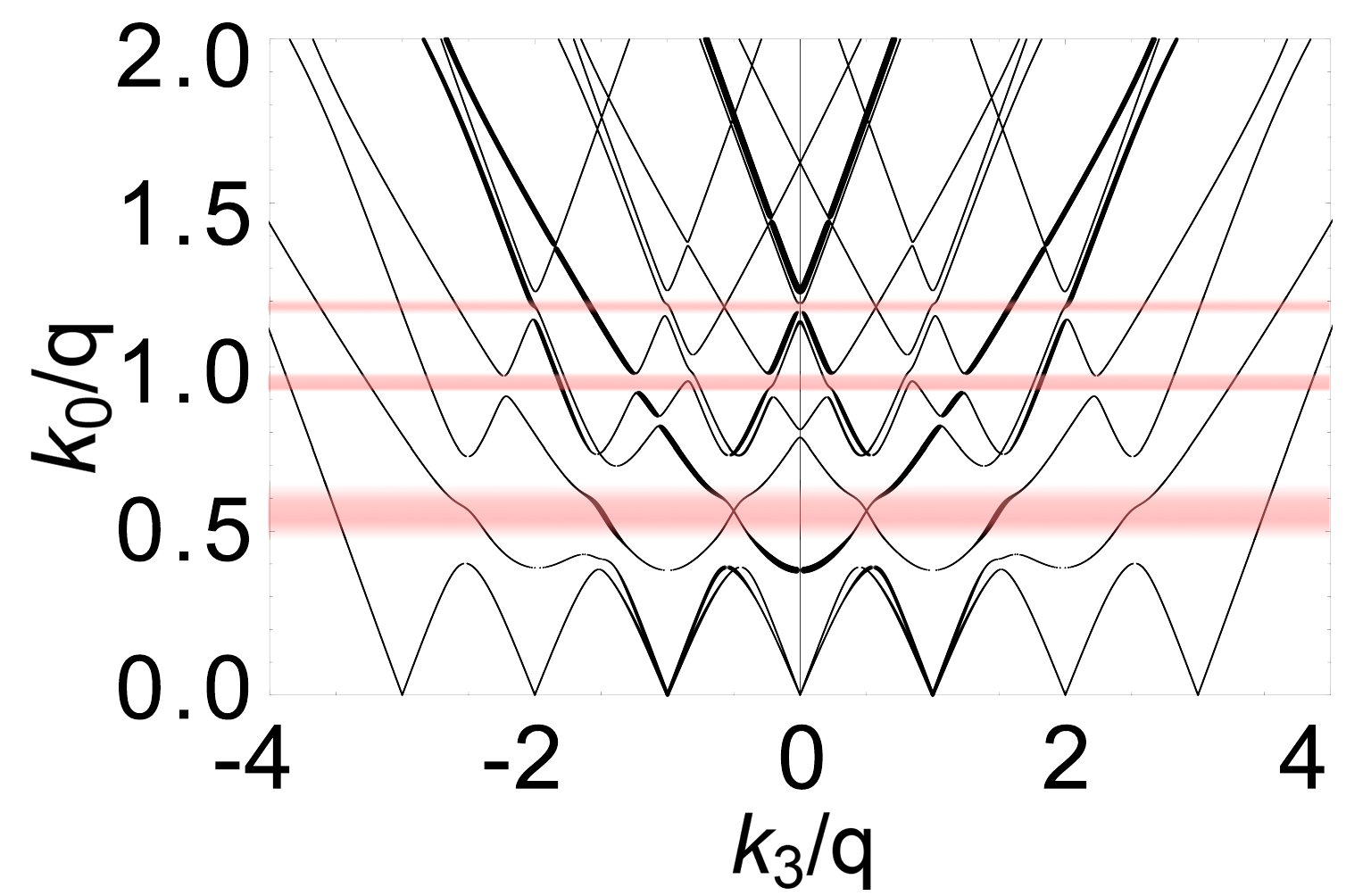}
	\caption{{\footnotesize The dispersion law in the $(2N+1)$-wave approximation with $N=2$ taking into account the root selection procedure. The size of the point corresponds to the root weight lying between $0$ and $1$. The pink bands show the chiral and complete band gaps. On the left panel: The dispersion law in the paraxial limit $n_\perp=0$. On the middle panel: $n_\perp=0.2$. On the right panel: The dispersion law in the nonparaxial regime, $n_\perp=0.5$. The other parameters are chosen as $\alpha=\pi/4$, $\e=\e_h=1$ , $v=1$, and $\omega_p=0.5q$.}}
	\label{ADisp}
\end{figure}
%%%%%%%%figure here

The trajectory of a particle with a charge $Ze$ moving uniformly and rectilinearly is given by
\begin{equation}
    \spx = \boldsymbol{\beta}t,\quad x^0 = t.
\end{equation}
Then the integral under the modulus sign in \eqref{GeneralProb} splits into several terms: the amplitude of edge radiation from the part of trajectory $z>0$:
\begin{equation}\label{EdgeRad1}
    -\frac{c}{\sqrt{2k_0V}} \frac{i\boldsymbol{\beta}\mathbf{f}(s,\spk)}{k_0(1-\spn\boldsymbol{\beta})};
\end{equation}
the amplitude of edge radiation from the part of trajectory $z<-L$:
\begin{equation}\label{EdgeRad2}
    \frac{c}{\sqrt{2k_0V}}\bigg( \frac{\boldsymbol{\beta}[d_+\mathbf{f}_{++}+d_-\mathbf{f}_{-+}]}{k_0(1-\spn\boldsymbol{\beta})} ie^{ik_0(1-\spn\boldsymbol{\beta})L/\beta_3} +\frac{\boldsymbol{\beta}[h_+\mathbf{f}_{+-}+h_-\mathbf{f}_{--}]}{k_0(1-\spn_\perp\boldsymbol{\beta}_\perp+n_3\beta_3)} ie^{ik_0(1-\spn_\perp\boldsymbol{\beta}_\perp+n_3\beta_3)L/\beta_3}\bigg);
\end{equation}
and the contribution to radiation amplitude from the periodic permittivity of the helical metamaterial in the region $-L \leq z\leq 0$:
\begin{multline}
    \frac{c}{\sqrt{2k_0V}}\sum_{i=1}^{6} b_i \sum_{l=-\infty}^\infty \Big[\frac{1}{2}\beta_-e^{i\vf}a_{+,i}^{l-1} + \frac{1}{2}\beta_+e^{-i\vf}a_{-,i}^{l+1}-\\
    -\beta_3 \big(\cos \alpha\frac{\omega_0\omega_p}{\sqrt{2}\bar{k}_3^2} \tilde{\Psi}_{i}^{l}  +\frac{k_\perp}{2\bar{k}_3^2}(k_3^i+ql)(a_{+,i}^{l-1} +a_{-,i}^{l+1} )\big)\Big] e^{-ik_3^i \vf /q}e^{-il\vf} g(x_i^l),
\end{multline}
where $a_{\pm,i}^l:=a_\pm^l(k_3^i)$ and
\begin{equation}\label{Harmonics}
\begin{gathered}
    x_i^l=k_0(1-\spn_\perp\boldsymbol{\beta}_\perp)-\beta_3(k_3^i+ql),\\
    g(x):=2\pi e^{i\frac{L}{2\beta_3}x}\delta_L(x),\qquad \delta_L(x):=\frac{\sin(\frac{L}{2\beta_3}x)}{\pi x}.
\end{gathered}
\end{equation}
In the case when the number of periods of the helical metamaterial is large, i.e., $L\gg \pi/|q|$, the modulus of the function $g(x)$, $x\in \mathbb{R}$, has a sharp maximum at $x=0$. Therefore, the main contribution to the radiation comes from the harmonics $x_i^l=0$. Notice that $|g(x+iy)|$ reaches the maximum at $x=0$ for fixed $y$ but in increasing $|y|$ this peak rapidly becomes blunt. This happens when $k_3^i(k_0)$ becomes complex-valued in the forbidden bands.

Thus the average number of radiated plane-wave photons \eqref{GeneralProb} is given by
\begin{equation}\label{ProbFinal}
\begin{split}
    dP(s,\spk)=& Z^2 e^2|c|^2 \bigg| \frac{\boldsymbol{\beta}[d_+\mathbf{f}_{++}+d_-\mathbf{f}_{-+}]}{k_0(1-\spn\boldsymbol{\beta})} ie^{ik_0(1-\spn\boldsymbol{\beta})L/\beta_3} +\frac{\boldsymbol{\beta}[h_+\mathbf{f}_{+-}+h_-\mathbf{f}_{--}]}{k_0(1-\spn_\perp\boldsymbol{\beta}_\perp+n_3\beta_3)} ie^{ik_0(1-\spn_\perp\boldsymbol{\beta}_\perp+n_3\beta_3)L/\beta_3}-\\
    &-\frac{i\boldsymbol{\beta}\mathbf{f}(s,\spk)}{k_0(1-\spn\boldsymbol{\beta})}+\sum_{i=1}^{6} b_i \sum_{l=-\infty}^\infty e^{i\frac{L}{2\beta_3}x_i^l} \delta_L(x_i^l)e^{-ik_3^i \vf /q}e^{-il\vf} \times\\
    &\times \Big[\frac{1}{2}\beta_-e^{i\vf}a_{+,i}^{l-1} + \frac{1}{2}\beta_+e^{-i\vf}a_{-,i}^{l+1} -\beta_3\big(\cos \alpha\frac{\omega_0\omega_p}{\sqrt{2}\bar{k}_3^2} \tilde{\Psi}_{i}^{l}  +\frac{k_\perp}{2\bar{k}_3^2}(k_3^i+ql)(a_{+,i}^{l-1} +a_{-,i}^{l+1} )\big)\Big] \bigg|^2  \frac{d\mathbf{k}}{4\pi k_0}.
\end{split}
\end{equation}
We will further use this general formula in numerical simulations and in theoretical analysis of the properties of transition radiation in a helical medium with strong spatial dispersion.

\section{$(2N+1)$-wave approximation}\label{NwaveApprox}

%%%%%%%%figure 2 here
\begin{figure}[tp]
	\centering
	\includegraphics*[width=0.45\linewidth]{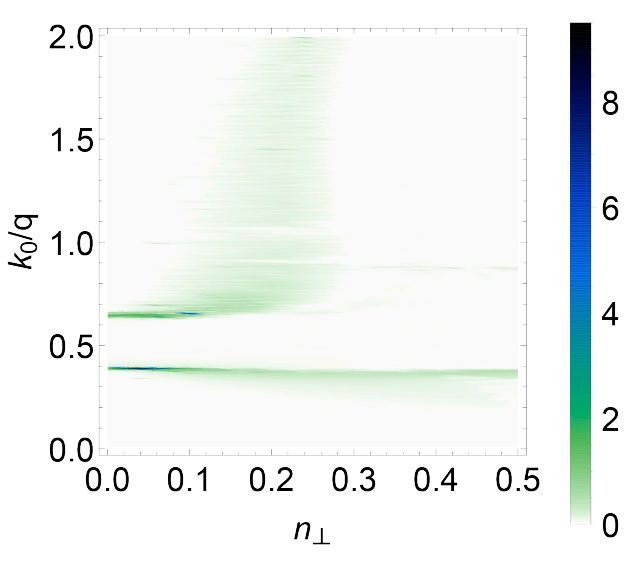}
	\includegraphics*[width=0.46\linewidth]{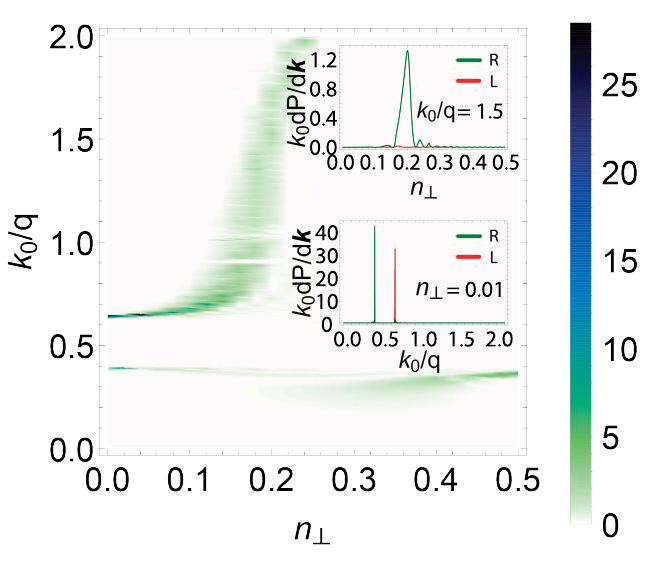}
	\caption{{\footnotesize The intensity of radiation of photons per volume of the cell in the momentum space from a relativistic electron with $\gamma=10$ moving along the symmetry axis of the helical medium with strong spatial dispersion as a function of the photon energy $k_0/q$ and the parameter $n_\perp=k_\perp/k_0$. The parameters of the medium are chosen as $\alpha=\pi/4$, $v=1$, $\e=\e_h=1$, and $\omega_p=0.5q$. On the left panel: the sample thickness is $L=40\pi/q$ corresponding to $20$ complete spiral periods. On the right panel: The sample thickness is $L=160\pi/q$ corresponding to $80$ complete spiral periods. The insets: The sections of the intensity plot with respect to $k_0/q$ and $n_\perp$ for the specimen width $L=160\pi/q$. The green line in the insets depicts the intensity of radiation of right-handed photons, whereas the red line indicates the intensity of radiation of left-handed photons.}}
	\label{A_DensIntensity}
\end{figure}
%%%%%%%%figure here

The infinite system of equations \eqref{MaxwellEq3} is useful in numerical simulations within the $(2N+1)$-wave approximation. In this case, the approximate solution to \eqref{MaxwellEq3} is sought in the form \eqref{FourierAnzatz}, where the summation is taken over a finite set of integers $l=\overline{-N,N} $ and the other coefficients of the Fourier series are set equal to zero. The system of equations obtained by these means from \eqref{MaxwellEq3} is already a finite system of homogeneous linear equations. The condition for the existence of non-trivial solutions to such a system (equality of the determinant to zero) leads to the dispersion relation which is a polynomial equation with respect to $k_3$ and $k_0$. In general, the order of the polynomial with respect to $k_3$ obtained in this way is $3\times2\times(2N+1)$, where $3$ is the number of independent fields in equation \eqref{MaxwellEq1}, $2$ corresponds to the two possible directions of propagation (to the left and to the right), and $2N+1$ is the number of non-zero selected blocks in the approximation. According to the fundamental theorem of algebra, there are $6(2N+1)$ solutions of this polynomial equation taking into account multiplicity. Hence the problem of selecting the main branches of the dispersion law arises and there should be six of them.

We follow the root selection procedure described in \cite{Vila2017}. Each solution $k_3^i(k_0)$, $i=\overline{1,6(2N+1)}$, corresponds to a subspace of eigenvectors $\{\mathbf{E}^{(s),i}\}$ of dimension $S$, $s=\overline{1,S}$, belonging to the null space of the matrix of the truncated equation \eqref{MaxwellEq3}. The number $S$ is the degeneracy of the eigenvalue $k_3^i(k_0)$. These eigenvectors are normalized to unity and then the weight is assigned to each such vector as
\begin{equation}
    w^{(s),i}=|E^{(s),i}_{3N+1}|^2+|E^{(s),i}_{3N+2}|^2+|E^{(s),i}_{3N+3}|^2,
\end{equation}
i.e., the squares of modules of the components of the eigenvector corresponding to the central $3\times3$ block of the coefficient matrix of the system of equations \eqref{MaxwellEq3} truncated in accordance with \eqref{FourierAnzatz} are summed. The root of the polynomial $k_3^i(k_0)$ is attributed by the weight
\begin{equation}
    w^i=\max_{s\in\{1,\ldots,S\}}(w^{(s),i}).
\end{equation}
Then the pairs $(w^i,k_3^i)$ are ordered in descending order by weight, and the first six values of $k_3^i(k_0)$ are selected as the main branches of the dispersion law. The branches of the dispersion law are shown in Fig. \ref{ADisp} for different $n_\perp$ taking into account the weighting within the root selection procedure in the $(2N+1)$-wave approximation with $N=2$. The greater the weight of the root, the larger the size of the corresponding point in the graph. In the paraxial limit, the system of equations \eqref{MaxwellEq3} is decoupled into $3\times3$ matrix blocks, and the principal branches of the dispersion law are selected naturally. This natural selection of the main branches agrees with the selection procedure described above. As $n_\perp$ increases, the blocks of the system of equations \eqref{MaxwellEq3} intertwine and the selection of the main branches of the dispersion law becomes a rather nontrivial issue.

For each main branch of the dispersion law $k_3^i(k_0)$ (now we assume $i=\overline{1,6}$), we obtain the solution $\{a_{\pm,i}^{l},\tilde{\Psi}_{i}^{l}\}$ of the truncated system of equations \eqref{MaxwellEq3}. As a result, we have an approximate general solution to the system of differential equations \eqref{MaxwellEq1} in the form \eqref{ModeMedium} but with a finite sum over $l=\overline{-N,N}$.

\section{Numerical simulations}\label{NumericalSimulation}

Let us consider the radiation of a relativistic electron, $Z=1$, traversing a helical metamaterial along the symmetry axis with velocity $\boldsymbol{\beta}=(0,0,\beta_3)$ and Lorentz factor $\gamma=10$. For the numerical analysis, we choose the parameters that qualitatively correspond to the actual helical metamaterials in the optical range as referenced in studies \cite{Morgado2012_1, Morgado2012_2,Morgado2016,Silverinha2008,Kashke2015,Kashke2016,Yang2012,Gansel2012}.
A typical value of the helix pitch in optical metamaterials is $p=2\pi/q\sim 1$ $\mu m$. However, we do not specify $q$ leaving it as a free parameter of the characteristic scale of energy and wave vector in a helical medium. For definiteness, we assume $q>0$. We take the remaining parameters of the model as follows: the helix tilt angle $\alpha=\pi/4$, the plasmon velocity $v=1$, the permittivity of the medium between thin conducting helices $\e_h=1$ coinciding with the permittivity $\e=1$ outside of the medium. For the plasma frequency, we consider the two values $\omega_p=0.5q$ and $\omega_p=1.3q$, which correspond to two different radiation regimes. The study of other regimes described in the paper \cite{KK2025} will be carried out in subsequent publications.

%%%%%%%%figure 3 here
\begin{figure}[tp]
\begin{minipage}[t]{0.53\linewidth}
	\centering
	\vspace{1pt} % для корректного выравнивания по верху
	\includegraphics*[width=\linewidth]{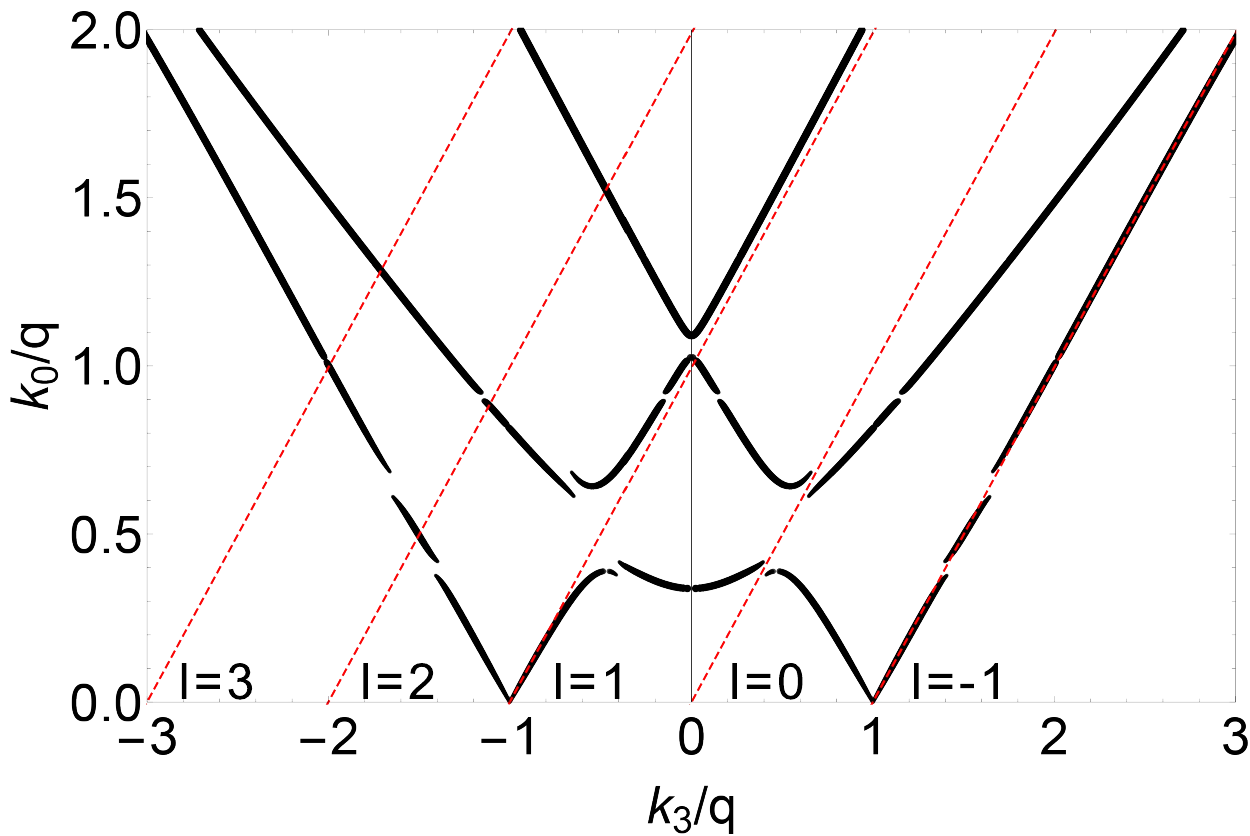}
\end{minipage}
%\\
%\hfill
\begin{minipage}[t]{0.47\linewidth}
	\centering
	\vspace{1pt} % для корректного выравнивания по верху	
\includegraphics*[width=0.45\linewidth]{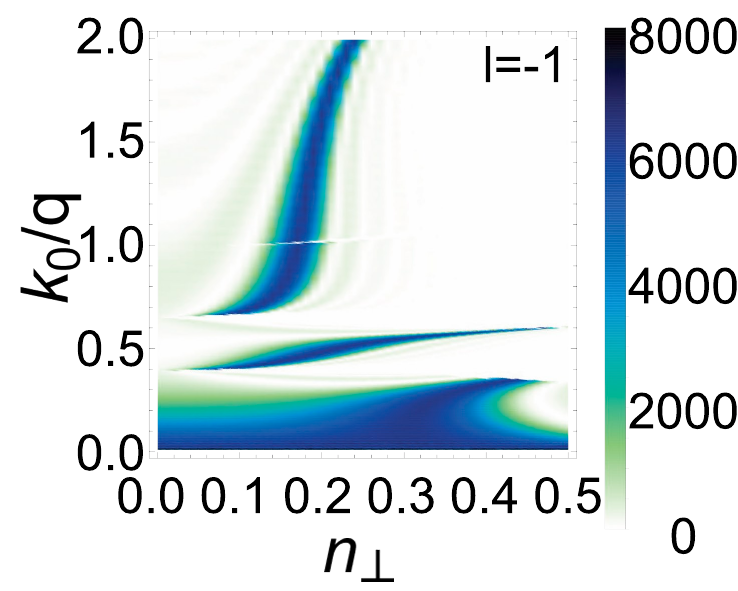}
\includegraphics*[width=0.45\linewidth]{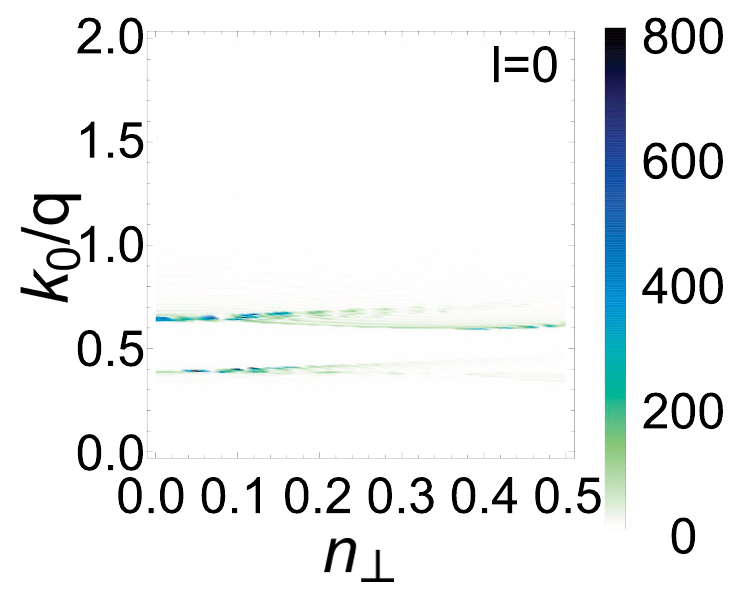}\\
\includegraphics*[width=0.45\linewidth]{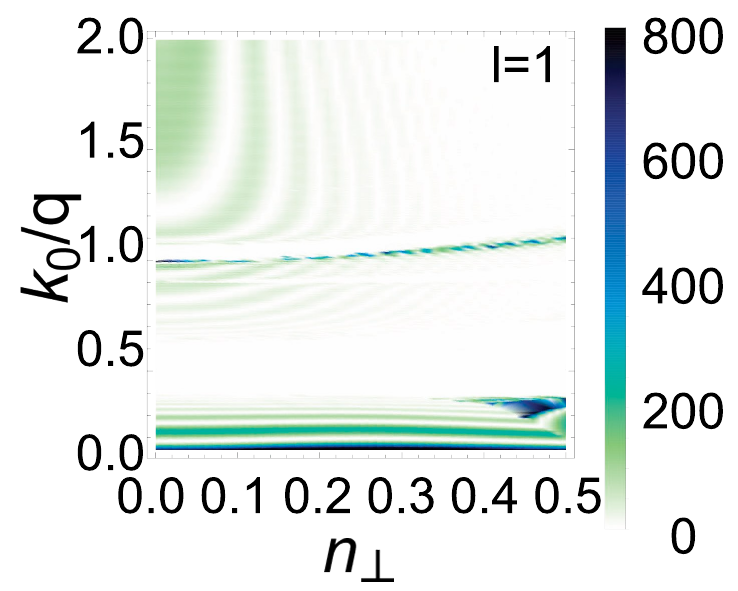}
\includegraphics*[width=0.45\linewidth]{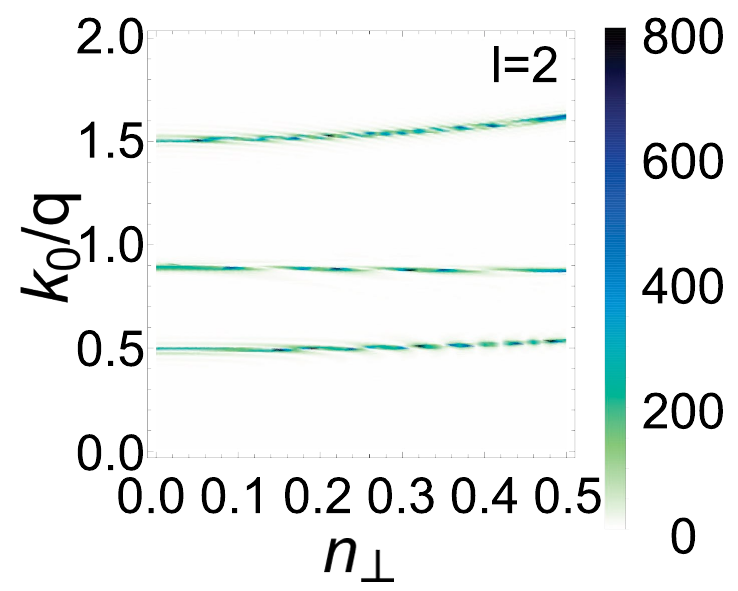}
\end{minipage}
	\caption{{\footnotesize On the left panel: The dispersion law with $n_\perp=0.2$ where the root selection procedure is applied. Only the real-valued roots are depicted. The red dashed lines indicate $k_0=\beta_3(k_3+ql)$, $l\in\mathbb{Z}$. The points of intersection of these lines with the dispersion law are the solutions to the equation $x_i^l=0$ and specify the positions of maxima of the radiation intensity. On the right panel: $|\sum_{i}\delta_L(x_i^l)|^2$ as a function of the photon energy $k_0/q$ and the parameter $n_\perp=k_\perp/k_0$. The electron Lorentz factor is $\gamma=10$. These plots determine the maxima of radiation with different $l$ produced by the electron moving along the symmetry axis of the helical medium. The parameters of the medium are chosen as $\alpha=\pi/4$, $v=1$, $\e=\e_h=1$, $\omega_p=0.5q$, and the specimen thickness is $L=160\pi/q$, i.e., $80$ complete spiral periods. It is seen that the red dashed line with $l=-1$ almost completely coincides with one of the branches of the dispersion law. This degenerate case corresponds to a wide maximum of radiation with $l=-1$ on the right panel and to a wide maximum of radiation visible in Fig. \ref{A_DensIntensity} for the radiation intensity. This contribution becomes prominent for sufficiently thick specimens.}}
	\label{A_TheorResonance}
\end{figure}
%%%%%%%%figure here

The results of numerical simulations for the radiation intensity of plane-wave photons per volume of the cell in the momentum space, $k_0dP(s,\spk)/d\spk$, as a function of photon energy $k_0/q$ and $n_\perp=k_\perp/k_0$ for different thicknesses of the helical metamaterial sample are shown in Fig. \ref{A_DensIntensity}. The numerical simulations were performed by using the general formula \eqref{ProbFinal}. In the paraxial regime $n_\perp \ll 1$, the sharp radiation maxima are visible corresponding approximately to the boundaries of the chiral band gap. With increasing $n_\perp$, the lower boundary of the chiral band gap is deformed insignificantly, which is also visible in the corresponding radiation maximum -- the lower branch depends weakly on $n_\perp$ (see Fig. \ref{A_DensIntensity}). At the same time, the branches of the dispersion law above the chiral band gap depend quite strongly on $n_\perp$ that results in a broadening of the upper branch of the emission maximum, which no longer coincides with the upper boundary of the chiral band gap. As the sample thickness increases, the radiation maxima become more pronounced. In the plots of sections in the insets in Fig. \ref{A_DensIntensity}, the green curve corresponds to the radiation intensity of right-handed circularly polarized photons and the red curve corresponds to the radiation intensity of left-handed circularly polarized photons.

%%%%%%%%figure 4 here
\begin{figure}[tp]
	\centering
	\includegraphics*[width=0.32\linewidth]{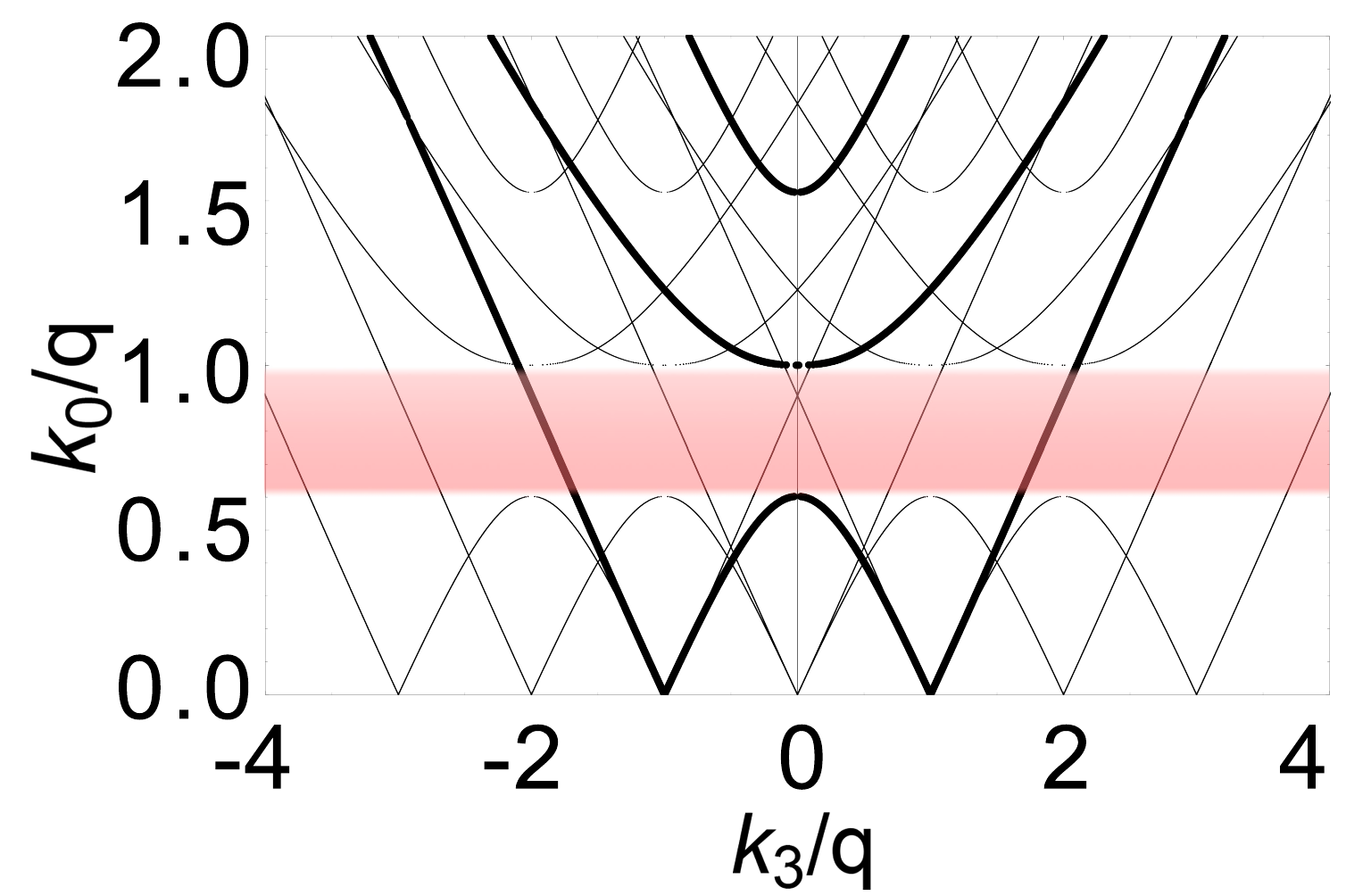}
	\includegraphics*[width=0.32\linewidth]{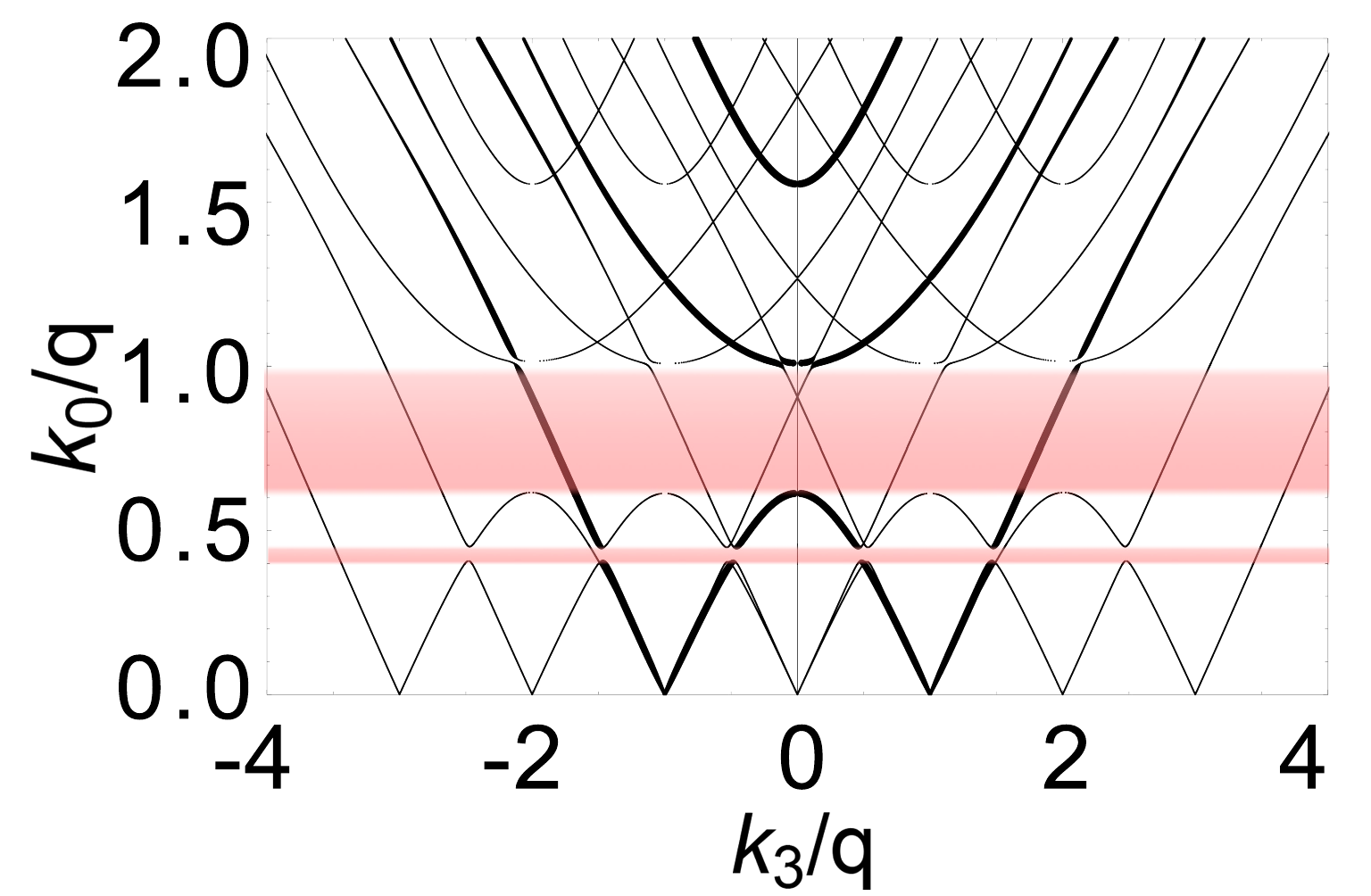}
	\includegraphics*[width=0.32\linewidth]{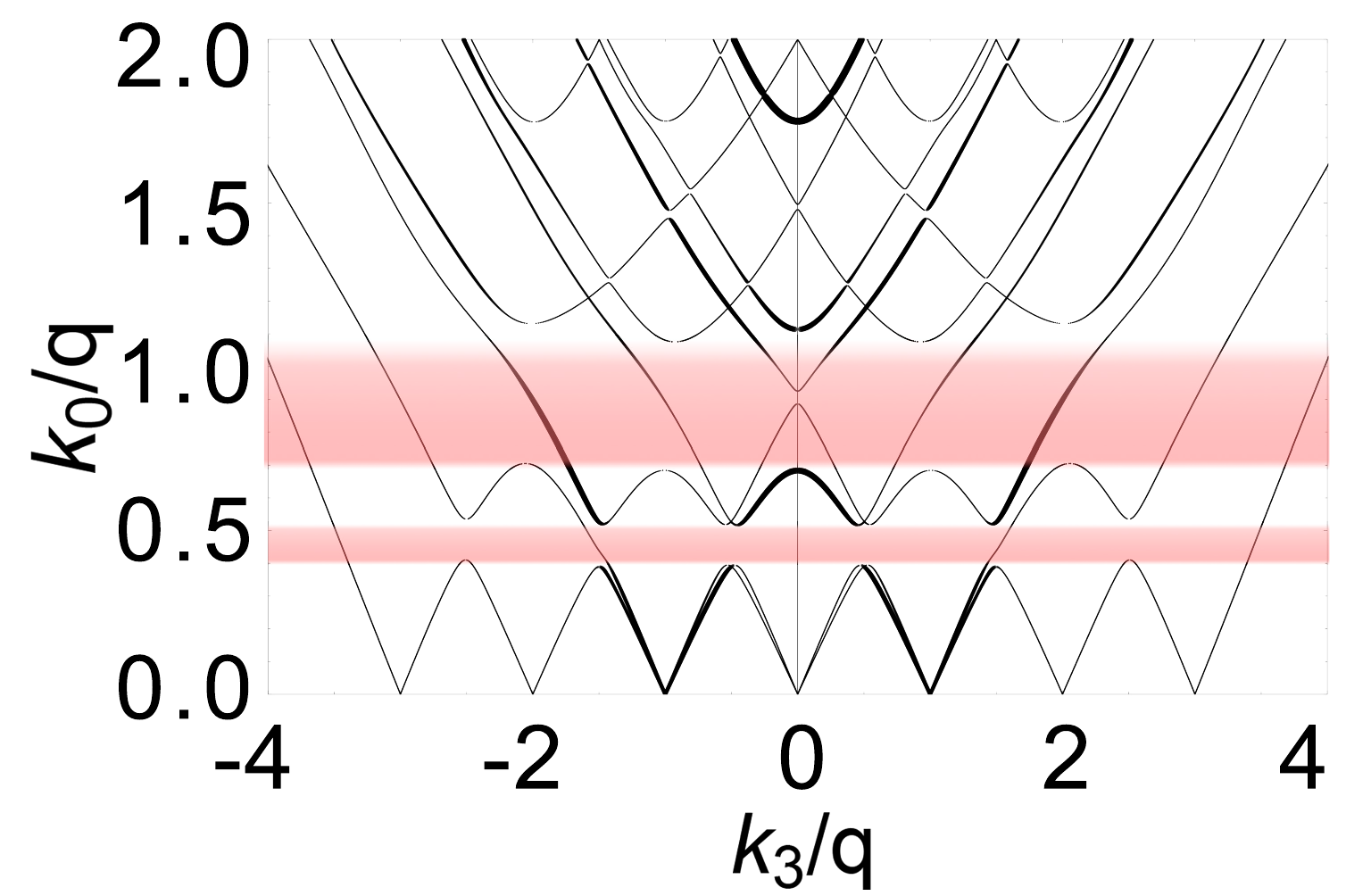}
	\caption{{\footnotesize The dispersion law in the $(2N+1)$-wave approximation with $N=2$ taking into account the root selection procedure. The size of the point corresponds to the root weight lying between $0$ and $1$. The pink bands show the chiral and complete band gaps. On the left panel: The dispersion law in the paraxial limit $n_\perp=0$ is shown. On the middle panel: $n_\perp=0.2$. On the right panel: The dispersion law in the nonparaxial regime, $n_\perp=0.5$, is presented. The other parameters are chosen as $\alpha=\pi/4$, $\e=\e_h=1$ , $v=1$, and $\omega_p=1.3q$. }}
	\label{BDisp}
\end{figure}
%%%%%%%%figure here

The locations of the intensity maxima can be found from the dispersion law of plasmon-polaritons in the medium. They are determined by the equation $x_i^l=0$ specifying the set of points where $|\de_L(x_i^l)|$ has a maximum. The equation $x_i^l=0$ can be conveniently solved graphically. The positions of the intensity maxima are realized at the intersection points of the plot of the dispersion law $k_0(k_3)$ with the family of straight lines (see the left panel in Fig. \ref{A_TheorResonance})
\begin{equation}\label{rad_cond}
    k_0=\beta_3 (k_3+ql), \qquad l\in \mathbb{Z}.
\end{equation}
In addition, the radiation polarization is determined by the polarization of the plasmon-polariton modes corresponding to the values of $k_3^i$ for which the radiation maxima are realized. The intersections of the plots at the points of the dispersion law where $\partial k^i_0/\partial k_3>0$ correspond to the main contribution to the radiation from a relativistic particle, which is recorded by the detector in the region $z\rightarrow+\infty$. The plasmon-polariton modes with $\partial k^i_0/\partial k_3<0$ propagate in the medium away from the detector and can contribute to the detected radiation only after reflection from the boundary $z=-L$. It is seen in Fig. \ref{A_TheorResonance} for the dispersion law at $n_\perp=0.2$ that the straight line with $l=0$ has two points of intersection with the dispersion law corresponding approximately to the locations of the boundaries of chiral band gap. These contributions are responsible for the two maxima in the radiation intensity at $n_\perp=0.2$. The straight line with $l=-1$ almost completely coincides with one of the branches of the dispersion law. This is, of course, a degenerate case due to a special choice of parameters. For a sufficiently thick sample, this contribution manifests itself as a broad maximum in the radiation intensity weakly depending on $k_0$ and overlapping with the upper maximum of the two intensity maxima corresponding to $l=0$ (see Fig. \ref{A_DensIntensity}).

The right panels in Fig. \ref{A_TheorResonance} represent the plots of $|\sum_{i}\delta_L(x_i^l)|^2$ as functions of $k_0/q$ and $n_\perp$ for different $l$. These plots allow us to decompose qualitatively the radiation intensity shown in Fig. \ref{A_DensIntensity} into the sum of contributions with different $l$, i.e., to obtain the radiation spectrum with respect to $l$. Indeed, comparing the plots on the right panels in Fig. \ref{A_TheorResonance} with the radiation intensity of plane-wave photons shown in Fig. \ref{A_DensIntensity}, we see that the terms with $l=-1$ and $l=0$ dominate in the radiation spectrum. This agrees with the above analysis of the radiation maxima based on the behavior of the dispersion law. This analysis can also be carried out in the opposite direction: starting from the experimental intensity (or probability) distribution of radiation for different $\be_3$, one can find the intersection points of the dispersion law of plasmon-polaritons $k_0(k_3)$ with the straight lines \eqref{rad_cond} and thereby reconstructs the dispersion law of plasmon-polaritons. Notice also that the results of the papers \cite{BKL2019,BKKL2021jml,BKKL2021pre} indicate that, in the medium with helical symmetry, the photons constituting the part of transition radiation corresponding to the number $l$ possess the projection of the total angular momentum onto the $z$ axis equal to $l$. We leave a detailed investigation of this property for future research.

\begin{figure}[tp]
	\centering
	\includegraphics*[width=0.45\linewidth]{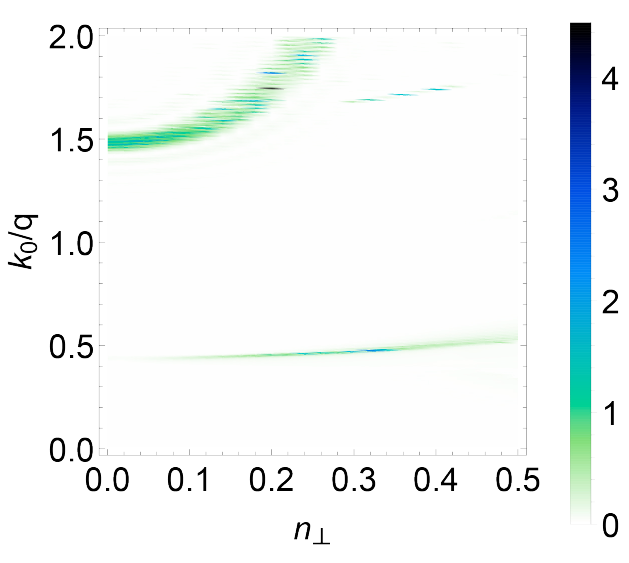}
	\includegraphics*[width=0.46\linewidth]{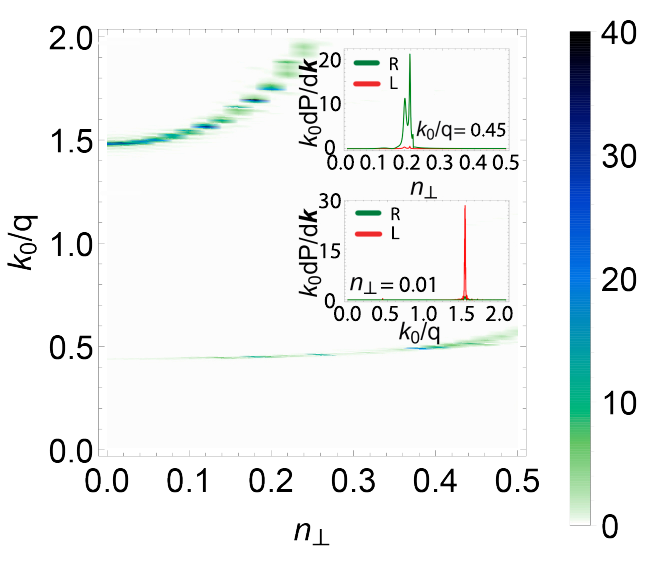}
	\caption{{\footnotesize The intensity of radiation of photons per volume of the cell in the momentum space from a relativistic electron with $\gamma=10$ moving along the symmetry axis of the helical medium with strong spatial dispersion as a function of the photon energy $k_0/q$ and the parameter $n_\perp=k_\perp/k_0$. The parameters of the medium are chosen as $\alpha=\pi/4$, $v=1$, $\e=\e_h=1$, and $\omega_p=1.3q$. On the left panel: The sample thickness is $L=40\pi/q$ corresponding to $20$ complete spiral periods. On the right panel: The sample thickness is $L=160\pi/q$ corresponding to $80$ complete spiral periods. The insets: The sections of the intensity plot with respect to $k_0/q$ and $n_\perp$ for the specimen width $L=160\pi/q$. The green line in the insets depicts the intensity of radiation of right-handed photons, whereas the red line indicates the intensity of radiation of left-handed photons.}}
	\label{B_DensIntensity}
\end{figure}

In Fig. \ref{BDisp}, the dispersion law is given for the same $n_\perp$ and other parameters as in Fig. \eqref{ADisp} but with different plasma frequency $\omega_p=1.3q$. In this case, the dispersion law has a shape different from \ref{ADisp} with a flat upper boundary of the chiral band gap. Notice also that for such parameters, as $n_\perp$ increases, a complete band gap appears below the chiral band gap. In Fig. \ref{B_DensIntensity}, the radiation intensities of plane-wave photons are presented as functions of energy $k_0/q$ and $n_\perp=k_\perp/k_0$ for different thicknesses of the helical metamaterial sample. The radiation spectra determined by $|\sum_{i}\delta_L(x_i^l)|^2$ are shown in Fig. \ref{B_TheorResonance}.

%%%%%%%%figure 5 here
%%%%%%%%figure here

In contrast to the regime with $\omega_p=0.5q$ considered above, in this case there is no branch of the dispersion law that would coincide with one of the straight lines \eqref{rad_cond} over a sufficiently large range of values of $k_0$. In this sense, the parameters chosen correspond to a general situation. Nevertheless, on the right panel in Fig. \ref{B_TheorResonance}, we see a similar wide maximum corresponding to $l=0$. Its width is explained by the fact that the red line \eqref{rad_cond} with $l=0$ is close to the tangent at the point of intersection with the dispersion law. This point belongs to the so-called ``plasmonic'' branch of the dispersion law (for details, see \cite{KK2025}). The corresponding maximum in radiation intensity arises as a result of the presence of a strong spatial dispersion in helical metamaterials and is absent, for example, in the radiation from relativistic charged particles traversing normally a cholesteric plate \cite{Shipov1991,Belyakov1972,Shipov1978,Velazquez2017,BKKL2021jml} or a plate of a homogeneous isotropic dielectric without strong spatial dispersion. This radiation is qualitatively distinct from the radiation concentrated at the maximum with $l=-1$ in Fig. \ref{A_DensIntensity}.

\section{Forward transition radiation}\label{ForwardRad}

The radiation coming from the plasmonic branch of the dispersion law is nonzero at $n_\perp=0$ which is not typical for transition and VC radiations in a medium without strong spatial dispersion. Indeed, consider the limit $n_\perp\rightarrow 0$ and $\beta_\perp=0$ in \eqref{ProbFinal}. The contributions corresponding to edge radiation vanish in this limit and
\begin{equation}\label{average_number_frwrd_rad}
	dP(s,\spk)=   Z^2 e^2  \omega_p^2  \beta^2_3\cos^2\alpha \frac{|c|^2}{\e_h} \bigg|\sum_{i=1}^{6} b_i
    \sum_{l\in\mathbb{Z}}e^{i\frac{L}{2\beta_3}x_i^l}
    \delta_L(x_i^l)e^{-ik_3^i \vf /q}e^{-il\vf}  \tilde{\Psi}_{i}^{l} \bigg|^2  \frac{d\mathbf{k}}{8\pi k^3_0}.
\end{equation}
We see that in this limit only the terms associated with the plasmon mode $\tilde{\Psi}(z)$ survive. In the absence of a strong spatial dispersion, for example, for cholesterics, such a contribution does not arise. The plasmon mode induces the emission of plane-wave photons along the axis of motion of a charged relativistic particle that coincides with the helical symmetry axis. For electrodynamics in a medium without spatial dispersion, the longitudinal component $A_3$ is excluded from the equations of motion and vanishes at $n_\perp=0$ (see \eqref{A_3} at $\Psi=0$ and Eq. (54) of \cite{BKKL2021jml}). As a result, it follows from the form of interaction of a point charged particle with an electromagnetic field that, for $n_\perp=0$, the charged particle moving along the $z$ axis ceases to interact with the quantum electromagnetic field and, consequently, to create radiation. In the presence of a strong spatial dispersion, the plasmonic field arises that according to \eqref{A_3} gives rise to $A_3\neq0$ even at $n_\perp=0$. This, in turn, leads to the possibility of forward radiation by a charged point particle moving along the $z$ axis in such a medium. This is one of the distinguishing features of the presence of a strong spatial dispersion.

%%%%%%%%figure 6 here
\begin{figure}[tp]
	\begin{minipage}[t]{0.53\linewidth}
		\centering
		\vspace{1pt} % для корректного выравнивания по верху
		\includegraphics*[width=\linewidth]{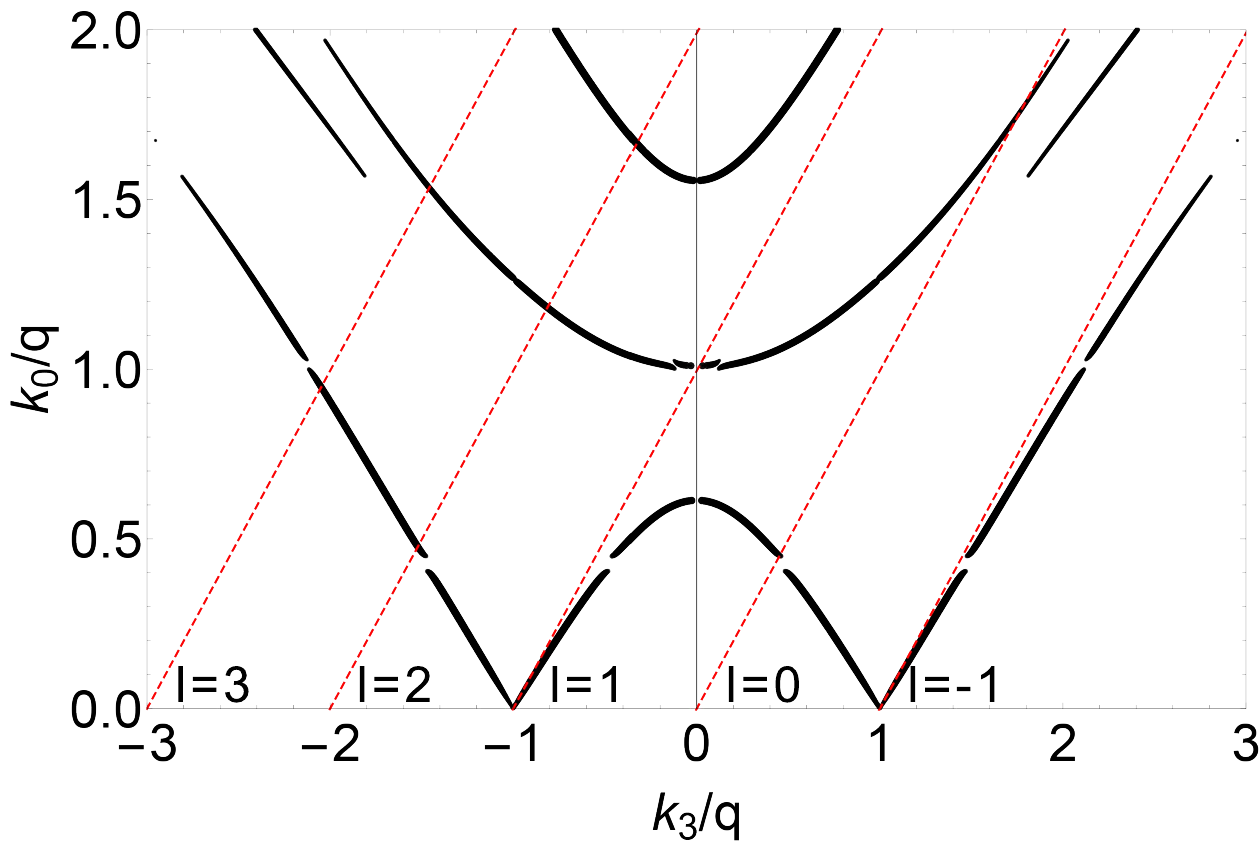}
	\end{minipage}
%	\hfill
	\begin{minipage}[t]{0.47\linewidth}
		\centering
		\vspace{1pt} % для корректного выравнивания по верху	
		\includegraphics*[width=0.45\linewidth]{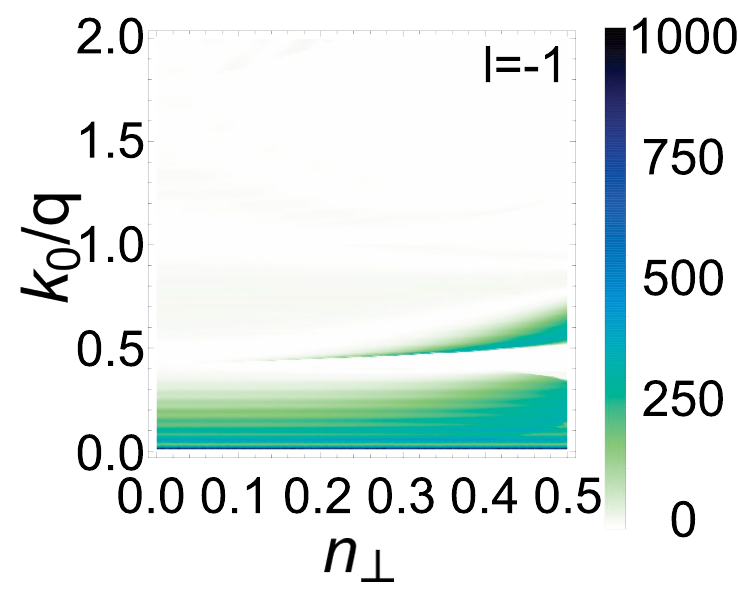}
		\includegraphics*[width=0.45\linewidth]{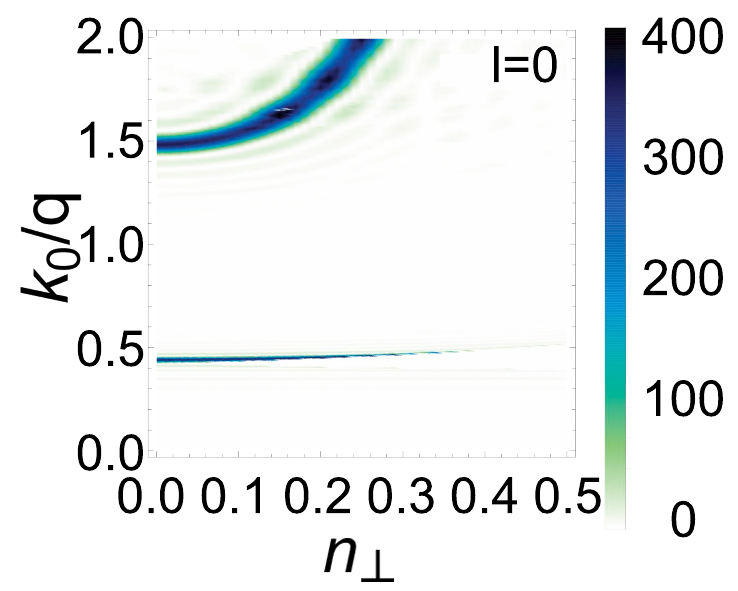}\\
		\includegraphics*[width=0.45\linewidth]{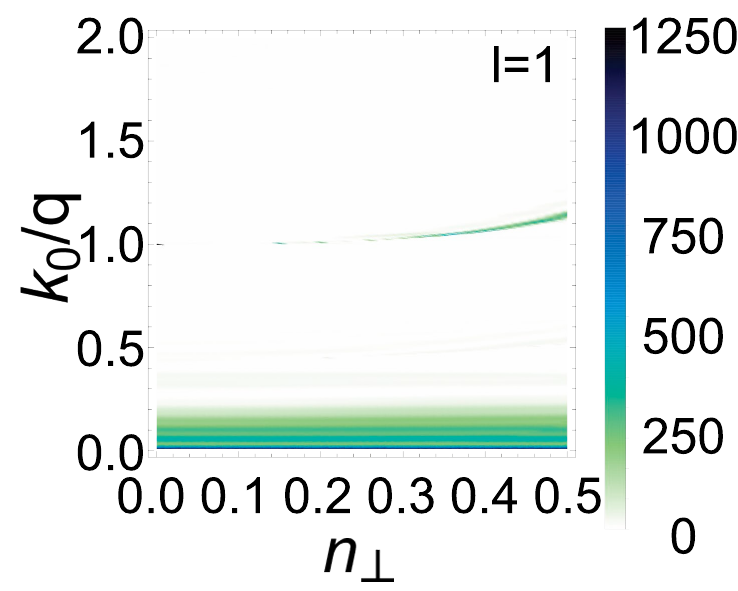}
		\includegraphics*[width=0.45\linewidth]{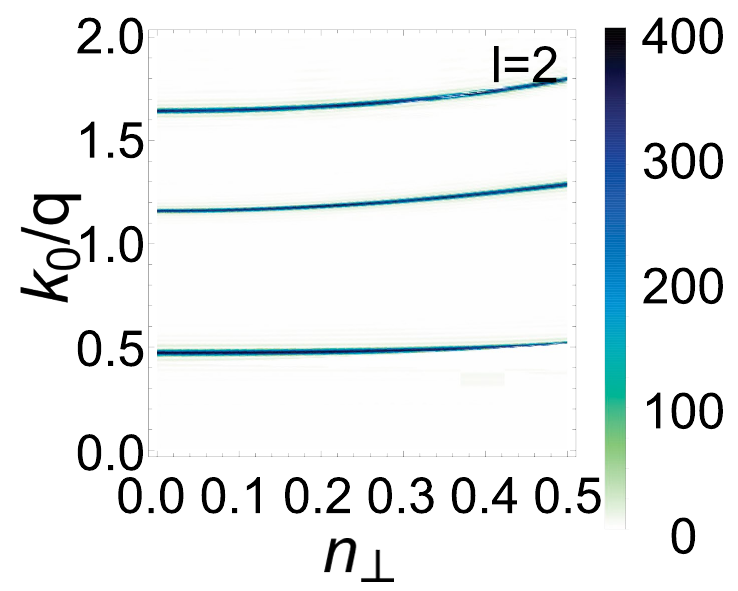}
	\end{minipage}
	\caption{{\footnotesize On the left panel: The dispersion law with $n_\perp=0.2$ where the root selection procedure is applied. Only the real-valued roots are depicted. The red dashed lines indicate $k_0=\beta_3(k_3+ql)$, $l\in\mathbb{Z}$. The points of intersection of these lines with the dispersion law are the solutions to the equation $x_i^l=0$ and specify the positions of maxima of the radiation intensity. On the right panel: $|\sum_{i}\delta_L(x_i^l)|^2$ as a function of the photon energy $k_0/q$ and the parameter $n_\perp=k_\perp/k_0$. The electron Lorentz factor is $\gamma=10$. These plots determine the maxima of radiation with different $l$ produced by the electron moving along the symmetry axis of the helical medium. The parameters of the medium are chosen as $\alpha=\pi/4$, $v=1$, $\e=\e_h=1$, $\omega_p=1.3q$, and the specimen thickness is $L=40\pi/q$, i.e., $20$ complete spiral periods. As against Fig. \ref{A_TheorResonance}, the red line does not coincide with one of the branches of the dispersion law. In this sense, the parameters chosen correspond to a general situation. Nevertheless, there is a wide maximum of intensity for $l=0$ on the right panel. It is wide since the red line is close to the tangent at the intersection point. The same maxima are visible in the radiation intensities presented in Fig. \ref{B_DensIntensity}. Notice that the contributions with $l=1,2,3$ for the photon energies we consider have the solutions of $x_i^l=0$ only for those values of $k_3$ where the group velocity $\partial k_0/\partial k_3<0$. Therefore, these contributions are suppressed in the radiation intensity given in Fig. \ref{B_DensIntensity} as these modes propagate away from the detector of photons and can be recorded by it only due to a weak reflection from the boundary $z=-L$. }}
	\label{B_TheorResonance}
\end{figure}
%%%%%%%%figure here

In the paper \cite{KK2025}, the exact expression for the dispersion law in the paraxial limit $n_\perp^2\ll1$ was found (see formula (39) of that paper). In the general case, it is rather cumbersome and reduces to a third-degree polynomial equation with respect to $k_0^2(k_3)$ or $k_3^2(k_0)$. On substituting $k_3$ into this expression, a sixth-degree polynomial equation on $k_0$ is obtained from \eqref{rad_cond}. Therefore, we will use an approximate expression for the dispersion law in the case when
\begin{equation}
    k_3^2+q^2\gg \omega_p^2\sin^2\al.
\end{equation}
When this condition is met, there are the three branches of the dispersion law (see Eq. (43) of \cite{KK2025})
\begin{equation}\label{disp_law_parax}
    k_0^2=(k_3\mp q)^2,\qquad k_0^2=\cos^2\al(\omega_p^2+k_3^2).
\end{equation}
The first two branches are called the right- and left-handed cholesteric branches for the upper and lower signs, respectively. The third branch of the dispersion law is called the polariton or plasmonic branch. As is shown in the paper \cite{KK2025}, the plasmonic branch describes the propagation of electromagnetic field perturbations along the conducting helical wires.

For the plasmon branch at $\be_3^2-\cos^2\al<0$, there are the two intersection points with the straight lines \eqref{rad_cond} for positive $k_0$ and fixed $l$,
\begin{equation}\label{max_intens_plasm}
    k_0=\frac{\be_3\cos\al}{\be_3^2-\cos^2\al} \Big[- q l\cos\al \pm \sqrt{\omega_p^2 (\be_3^2-\cos^2\al) +\be_3^2q^2l^2}\Big],
\end{equation}
provided
\begin{equation}\label{rad_cond_ordinar_plasm}
    0<\cos^2\al -\be_3^2\leqslant \frac{\be_3^2q^2l^2}{\omega_p^2},\qquad ql>0.
\end{equation}
From the last condition, in particular, it follows that $k_0>\be_3 k_3$ in the case under consideration.

\begin{figure}[tp]
	\centering
	\includegraphics*[width=0.49\linewidth]{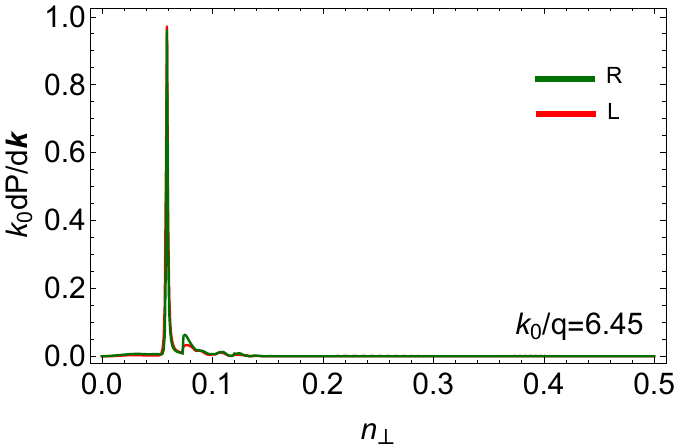}
	\includegraphics*[width=0.49\linewidth]{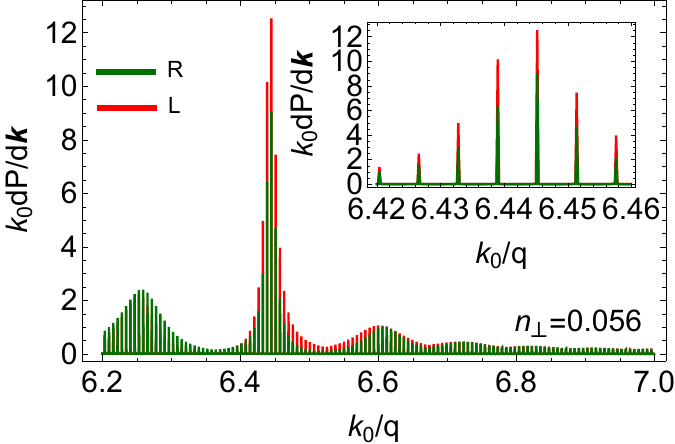}
	\caption{{\footnotesize	The intensity of VC radiation with $l=0$ produced by a relativistic electron moving along the symmetry axis of the helical medium with the Lorentz factor $\gamma=10$ as a functions of $n_\perp=k_\perp/k_0$ for the fixed energy $k_0=6.6q$ and as a function of the photon energy $k_0/q$ for the fixed $n_\perp=0.056$. The green line depicts the intensity of radiation of right-handed photons, whereas the red line indicates the intensity of radiation of left-handed photons. The parameters of the medium are chosen as  $\alpha=0.04\pi$, $v=1$, $\e=\e_h=1$ $\omega_p=0.5q$, and $L=160\pi/q$ corresponding to $80$ complete spiral periods. The inset: The fine structure of the radiation intensity.}}
	\label{C_Cherenkov}
\end{figure}

For higher electron velocities, $\be_3^2-\cos^2\al>0$, there is only one intersection point of the plasmon branch of the dispersion law with the straight lines \eqref{rad_cond} for positive $k_0$ and fixed $l$. It is described by expression \eqref{max_intens_plasm} where the plus sign is taken. The conditions \eqref{rad_cond_ordinar_plasm} are absent in this case. For $ql<0$ we have $k_0<\be_3 k_3$, i.e., this radiation is created inside the Cherenkov cone and the anomalous Doppler effect is realized for it \cite{Ginzburg1979,Nezlin1976} provided $\partial k_0/\partial k_3>0$ at this point. VC radiation corresponds to $l=0$ \cite{TerMikael1969,Ginzburg1979,Ginzburg1990} and we have
\begin{equation}
    k_0=\frac{\omega_p\be_3\cos\al}{\sqrt{\be_3^2-\cos^2\al}},\qquad k_0=\be_3k_3.
\end{equation}
Obviously, it is formed only for $\be_3^2-\cos^2\al>0$. The radiation intensity in the vicinity of the Cherenkov maximum is shown in Fig. \ref{C_Cherenkov}. It has a distinctive fine structure in the form of a periodic set of peaks. The position of these resonances can be explained in terms of the properties of the plasmon field $\Psi$. As can be seen from the Maxwell equations \eqref{MaxwellEq0} and from the detailed analysis given in Section 4 of the paper \cite{KK2025}, the plasmon field propagates along the conducting spiral wires with the velocity $v=1$ and satisfies the zero boundary conditions. Consequently, there are resonances at
\begin{equation}
    k_0 l_h=\pi n,\qquad n\in \mathbb{Z},
\end{equation}
where $l_h$ is the length of the conducting helix, which is related to the plate thickness as $l_h=L/\cos\al$. As a result, we obtain the condition for resonances
\begin{equation}\label{plasmon_reson}
    k_0=\cos\al \frac{\pi n}{L}.
\end{equation}
Numerical analysis shows that this expression describes with good accuracy the distance between peaks in the fine structure of the radiation intensity.

%%%%%%%%figure 7 here
%%%%%%%%figure here

A similar analysis for the emission maxima coming from the cholesteric branches of the dispersion law leads to two points of the radiation maxima for each branch
\begin{equation}\label{rad_cond_chol}
    k_0=\frac{\be_3q(l\pm 1)}{1\pm \be_3},\qquad q(l\pm 1)\geqslant0.
\end{equation}
where the sign in the numerator and the sign in the last inequality agree with the sign in the dispersion law for the cholesteric branch \eqref{disp_law_parax}. The sign in the denominator of the fraction in \eqref{rad_cond_chol} can be arbitrary. In this case, the anomalous Doppler effect is realized at $l=-\sgn(q)$ for the right (with $q>0$) or left (with $q<0$) cholesteric branch of the dispersion law. Formally, in this case $k_0=0$. However, as the numerical simulations reveal (see Fig. \ref{A_TheorResonance}, \ref{B_TheorResonance}), the corresponding straight line \eqref{rad_cond} approximates closely the dispersion law in the region of small $k_0$ and can remain close to this branch over a significant range of $k_0$ values. The numerical analysis also shows that the strongest contribution to the radiation comes from VC radiation, $l=0$, and from the region where the anomalous Doppler effect is realized, $l=-1$ for $q>0$. Notice that the radiation from the right cholesteric branch has predominantly a right-handed circular polarization, while the radiation from the left cholesteric branch exhibits predominantly a left-handed circular polarization.

\section{Conclusion}

Let us sum up the results. In the framework of the effective model of the helical metamaterial with strong spatial dispersion \eqref{ModelKernel}, we have described the properties of transition radiation produced by a relativistic charged particle moving with constant velocity through the plate made of this metamaterial along its symmetry axis. We suppose that this plate is positioned orthogonally to the symmetry axis. As expected \cite{TerMikael1969,Bazylev1987,Ginzburg1990,Shipov1991,Shipov1978,Velazquez2017}, a periodicity of the dielectric permittivity of the metamaterial along the symmetry axis gives rise to the presence of the Bragg maxima in the intensity of transition radiation which positions in the spectrum domain are determined by the dispersion law of plasmon-polaritons in the metamaterial. We have developed the procedure allowing one to restore the dispersion law of plasmon-polaritons given the positions of the transition radiation maxima in momentum space for different velocities of a charged particle (see Figs. \ref{A_TheorResonance}, \ref{B_TheorResonance} and comments to them in Sec. \ref{NumericalSimulation}). Inversely, this procedure with the knowledge of the dispersion law of plasmon-polariton modes in the metamaterial and their polarization properties makes it possible to predict the locations of the intensity maxima of transition radiation and its polarization.

Using the results of the paper \cite{KK2025}, where the dispersion law of plasmon-polaritons in the helical metamaterial with strong spatial dispersion was thoroughly investigated, in Sec. \ref{NumericalSimulation} we have simulated numerically transition radiation for the two typical regimes of the dispersion law with $\omega_p=0.5q$ and $\omega_p=1.3q$. These values of the plasma frequency correspond qualitatively to the existing helical metamaterials in the optical spectral range \cite{Morgado2016,Kashke2016,Kashke2015,Morgado2012_2,Morgado2012_1,Silverinha2008} with the helix pitch $p=2\pi/q\sim 1$ $\mu$m. However, we have represented the results in the form keeping the parameter $q$ as a characteristic energy scale that enables one to apply the obtained results out of the optical spectral domain. For the plasma frequency $\omega_p=0.5q$, the dispersion law of plasmon-polaritons has a typical form of the dispersion laws with $\omega_p\lesssim q$. For the plasma frequency $\omega_p=1.3q$, one of the branches of the dispersion law is flattened out, $k_0\sim k_3^4$, near the upper boundary of the chiral band gap for small $k_3$ and $n_\perp$. In the former case, the sharp maxima corresponding to the lower and upper boundaries of the chiral band gap are clearly seen in Fig. \ref{A_DensIntensity} for the radiation intensity in the paraxial regime. In the latter case, only the contribution of the lower boundary of the chiral band gap is seen in Fig. \ref{B_DensIntensity} for small $n_\perp$. Besides, for both cases there are the similar wide maxima of radiation above the chiral forbidden band possessing, however, essentially different interpretations. For the case $\omega_p=0.5q$, this maximum is caused by a peculiar behavior of the dispersion law such that the charged particle excites a considerable part of the cholesteric branch of the dispersion law of plasmon-polaritons in the metamaterial. In the case $\omega_p=1.3q$, a similar maximum is generated due to excitation of the plasmonic branch of the dispersion law by a charged particle. The polarization of transition radiation is determined by the polarization of the excited plasmon-polariton branches of the dispersion law. In particular, the radiation at the wide maximum in Fig. \ref{A_DensIntensity} for $n_\perp\approx 0.2$ caused by excitation of the cholesteric branch of the dispersion law has a right-handed circular polarization. The forward radiation for the parameters chosen in Figs. \ref{A_DensIntensity}, \ref{B_DensIntensity} has a circular polarization when it is caused by excitation of the cholesteric branches of the plasmon-polariton dispersion law.

It turns out that the strong spatial dispersion resulting in appearance of an additional degree of freedom -- the plasmon field, changes qualitatively the properties of transition radiation for small radiation angles, $n_\perp \ll 1$. The presence of the plasmon field hybridized with the electromagnetic modes leads to new emission channels inaccessible in conventional chiral or isotropic dielectrics. In particular, in contrast to transition radiation generated by a charge propagating in a medium without strong spatial dispersion such as dielectric or cholesteric plates, the intensity of transition radiation produced by a charge moving in a plate made of the helical metamaterial with strong spatial dispersion differs from zero even for $n_\perp=0$, i.e., there is the forward radiation in this case. The intensity of this radiation is described by expression \eqref{average_number_frwrd_rad} multiplied by the photon energy $k_0$. We have shown that, for $\be_3>\cos\al$, classical VC radiation corresponding to the zeroth Bragg maximum, $l=0$, is generated in the metamaterial. In this case, the anomalous Doppler effect occurs for the Bragg maxima in the intensity of radiation with $ql<0$. The numerical simulations have revealed that, for $\be_3>\cos\al$, VC radiation, $l=0$, and the radiation where the anomalous Doppler effect is realized, $l=-\sgn(q)$, give the main contribution to the intensity of radiation. For the parameters of the metamaterial we have considered, VC radiation is generated by the plasmonic branch of the dispersion law and its intensity possesses a specific fine structure (see Fig. \ref{C_Cherenkov}) arising due to the presence of the plasmonic resonances \eqref{plasmon_reson} in the plate made of conducting wires. These properties unambiguously indicate the existence of a strong spatial dispersion and an additional degree of freedom associated with it in the metamaterial.

Thus we see that the helical metamaterials with strong spatial dispersion provide a means for bright sources of transition radiation with the energy spectrum and the polarization governed by the parameters of this metamaterial. The availability of a rather simple effective model describing the electromagnetic properties of the metamaterial makes it possible not only to accelerate numerical simulations of scattering and emission of electromagnetic waves in this medium but also to derive simple estimates for the positions of maxima of radiation and its polarization. The theory developed in the present paper allows one to solve also the inverse problem, viz., to recover the spectrum of plasmon-polariton modes in metamaterials with the aid of the properties of transition radiation measured in experiments.

\paragraph{Acknowledgments.}

The study was supported by the Russian Science Foundation, grant No. 25-21-00283, https://rscf.ru/en/project/25-21-00283/

%\newpage

%\begin{center}

%\end{center}

 \end{document}